\documentclass[12pt]{article}
\usepackage{citesort,fullpage,epsfig,psfrag,graphics,amsbsy,amssymb}
\usepackage{caption}
\newcommand{\beq}{\begin{equation}}
\newcommand{\eeq}{\end{equation}}
\newcommand{\bea}{\begin{eqnarray}}
\newcommand{\eea}{\end{eqnarray}}
\newcommand{\bfs}{\boldsymbol}

\newcommand{\Tr}{{\rm Tr}}
\newcommand{\be}{\begin{equation}}
\newcommand{\ee}{\end{equation}}
\newcommand{\bq}{\begin{eqnarray}}
\newcommand{\eq}{\end{eqnarray}}

\def\math{\mathsurround=0pt }
\def\leftrightarrowfill{$\math \mathord\leftarrow \mkern-6mu 
 \cleaders\hbox{$\mkern-2mu \mathord- \mkern-2mu$}\hfill
 \mkern-6mu \mathord\rightarrow$}
\def\overleftrightarrow#1{\vbox{\ialign{##\crcr
     \leftrightarrowfill\crcr\noalign{\kern-1pt\nointerlineskip}
     $\hfil\displaystyle{#1}\hfil$\crcr}}}

\let\l=\lambda

 \def\bd{\begin{document}} \def\ed{\end{document}}
\def\ds{\documentstyle} \let\fr=\frac \let\bl=\bigl \let\br=\bigr
\let\Br=\Bigr \let\Bl=\Bigl
\let\bm=\bibitem
\let\na=\nabla
\let\pa=\partial \let\ov=\overline
\def\ft#1#2{{\textstyle{{\scriptstyle #1}\over {\scriptstyle #2}}}}
\def\fft#1#2{{#1 \over #2}}
\def\vp{\varphi}
\def\sst#1{{\scriptscriptstyle #1}}
\def\oneone{\rlap 1\mkern4mu{\rm l}}
\def\td{\tilde}
\def\wtd{\widetilde}
\def\dalemb#1#2{{\vbox{\hrule height .#2pt
        \hbox{\vrule width.#2pt height#1pt \kern#1pt
                \vrule width.#2pt}
        \hrule height.#2pt}}}
\def\square{\mathord{\dalemb{6.8}{7}\hbox{\hskip1pt}}}
\def\wtd{\widetilde}
\def\R{\rlap{\rm I}\mkern3mu{\rm R}}
\def\im{{\rm i}}
\def\tilg{\tilde{g}}
\def\tilF{\tilde{F}}
\def\tilA{\tilde{A}}
\def\varf{\varphi}
\def\tilf{\tilde{\phi}}
\def\tilh{\tilde{h}}
\def\rme{{\rm e}}
\def\ep{\epsilon}
\def\0{{(0)}}
\def\9{{(9)}}
\def\8{{(8)}}
\def\7{{(7)}}
\def\6{{(6)}}
\def\5{{(5)}}
\def\4{{(4)}}
\def\3{{(3)}}
\def\2{{(2)}}
\def\1{{(1)}}
\newcommand{\trace}{{\rm Tr}}
\newcommand{\ub}{\overline{U}}
\newcommand{\vb}{\overline{V}}
\newcommand{\uh}{\widehat{U}}
\newcommand{\vh}{\widehat{V}}
\newcommand{\ubh}{\overline{\widehat{U}}}
\newcommand{\vbh}{\overline{\widehat{V}}}
\newcommand{\lb}{\bar{\l}}
\newcommand{\Fb}{\overline{F}}
\newcommand{\Fh}{\widehat{F}}
\newcommand{\Fbh}{\overline{\widehat{F}}}
\newcommand{\Ab}{\overline{A}}
\newcommand{\Ah}{\widehat{A}}
\newcommand{\Abh}{\overline{\widehat{A}}}
\newcommand{\Gb}{\overline{G}}
\newcommand{\Gh}{\widehat{G}}
\newcommand{\Gbh}{\overline{\widehat{G}}}
\newcommand{\Pb}{\overline{P}}
\newcommand{\Ph}{\widehat{P}}
\newcommand{\Pbh}{\overline{\widehat{P}}}
\newcommand{\Qb}{\overline{Q}}
\newcommand{\Qh}{\widehat{Q}}
\newcommand{\Qbh}{\overline{\widehat{Q}}}
\newcommand{\Bb}{\overline{B}}
\newcommand{\Bh}{\widehat{B}}
\newcommand{\Bbh}{\overline{\widehat{B}}}
\newcommand{\fhns}{\hat{F}^{\rm (NS)}}
\newcommand{\fhrr}{\hat{F}^{\rm (RR)}}
\newcommand{\ahns}{\hat{A}^{\rm (NS)}}
\newcommand{\ahrr}{\hat{A}^{\rm (RR)}}
\newcommand{\hhrr}{\hat{H}^{\rm (RR)}}
\newcommand{\hchi}{\hat{\chi}}
\newcommand{\hphi}{\hat{\phi}}
\newcommand{\htau}{\hat{\tau}}
\newcommand{\cG}{{\cal G}}
\newcommand{\cGb}{\overline{{\cal G}}}
\newcommand{\cH}{{\cal H}}
\newcommand{\cP}{{\cal P}}
\newcommand{\cPb}{\overline{{\cal P}}}
\newcommand{\cQ}{{\cal Q}}
\newcommand{\cQb}{\overline{{\cal Q}}}
\newcommand{\cM}{{\cal M}}
\newcommand{\cN}{{\cal N}}
\newcommand{\cO}{{\cal O}}
\newcommand{\cD}{{\cal D}}
\newcommand{\cL}{{\cal L}}

\newcommand{\vpp}{\mbox{$\langle{\scriptstyle++}\rangle$}}
\newcommand{\vmp}{\mbox{$\langle{\scriptstyle-+}\rangle$}}
\newcommand{\vppp}{\mbox{$\langle{\scriptstyle+++}\rangle$}}
\newcommand{\vmpp}{\mbox{$\langle{\scriptstyle-++}\rangle$}}
\newcommand{\vpmp}{\mbox{$\langle{\scriptstyle+-+}\rangle$}}
\newcommand{\goesas}[1]{{}_{{\displaystyle\sim}\atop#1}}
\newcommand{\impliesas}[1]{{}_{{\displaystyle{\Longrightarrow}}\atop#1}}
\renewcommand{\thepage}{\arabic{page}}

\begin{document}
\setlength{\captionmargin}{36pt}
\begin{titlepage}
\begin{flushright}
\phantom{UFIFT}
\end{flushright}

\vskip 3cm
\begin{center}
\begin{large}
{\bf Determinants for the Lightcone Worldsheet}
\footnote{Supported in part by the Department
of Energy under Grant No. DE-FG02-97ER-41029.} 
\end{large}
\vskip 2cm
{\large 
Charles B. Thorn
\footnote{E-mail  address: {\tt thorn@phys.ufl.edu}}
}
\vskip0.20cm
{\it Institute for Fundamental Theory,\\
Department of Physics, 
University of Florida
Gainesville FL 32611
}

\vskip24pt
\end{center}
\begin{abstract}
\noindent The evaluation of the determinant of the
Laplacian defined on two dimensional regions of various shapes
is an essential ingredient in calculating the scattering
amplitudes of strings. In lightcone parameterization the
regions are rectangular in shape with several slits 
of different length and location cut parallel to the
$\tau$ axis of the rectangle. This paper offers a compendium
of applications of the methods of Kac and McKean and Singer
to the calculation of such worldsheet determinants. 
Particular attention is paid to the effect of corners on
the determinants. The effect of corners joining edges 
with like boundary conditions is implicit in Kac's results.
We discuss the generalization to a corner joining a 
Dirichlet edge to a Neumann edge, and apply it to
a scattering amplitude involving D-branes.
\end{abstract}

\vfill
\end{titlepage}
\section{Introduction}
\begin{figure}[ht]
\begin{center}
\includegraphics[width=2.9in,height=1.7in]{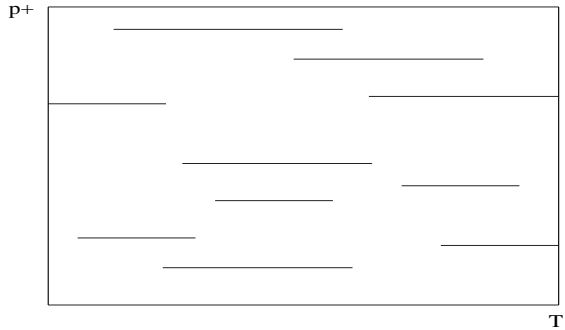}\\
\caption{Mandelstam Interacting String Diagram}
\label{lcworldsheet}
\end{center}
\end{figure}

The lightcone quantization of string \cite{goddardrt,goddardgrt}
was employed by Mandelstam \cite{mandelstamlc,mandelstamnsr}
to describe interacting string theory via the sum over path histories in which
interactions between strings are interpreted simply as breaking and
joining processes as depicted in Fig.~\ref{lcworldsheet}.
The lightcone worldsheet is parameterized by taking the
evolution parameter $\tau$ to be $ix^+$, where the $i$ reflects a
Wick rotation to imaginary $x^+$; 
and by labelling points on the string
by a parameter $\sigma$ defined so that density of $P^+$ momentum is unity.
Then the dimensions of the world sheet are 
\bea
T=ix^+=i(t+z)/\sqrt2,\qquad P^+=(p^0+p^z)/\sqrt2 ,
\eea
where $x^\mu$ and $p^\mu$ are the spacetime coordinates and
total four momentum of the string.
The diagram in Fig.~\ref{lcworldsheet} 
describes the time evolution of a system of open strings,
breaking and rejoining as shown by the horizontal lines.

For the {critical} open bosonic string (i.e. the spacetime
dimension $D=26$), the worldsheet 
path history integrates over the transverse coordinates 
${\bfs x}(\sigma,\tau)$
and uses the lightcone action for the free open string:
\bea
S_{l.c.}={1\over2}\int_0^Td\tau\int_0^{P^+}d\sigma\left[\left({\partial{\bfs
x}\over\partial\tau}\right)^2+T_0^2\left({\partial{\bfs
x}\over\partial\sigma}\right)^2\right]
\eea
The transverse coordinates, defined on the domain of the lightcone diagram,
are discontinuous across horizontal lines. For each beginning and end 
of a horizontal line there is a factor of string coupling 
$g$. Then the sum over all planar open string loops
is simply the sum over the number, lengths and locations of those
horizontal lines. 

\begin{figure}[ht]
\begin{center}
\includegraphics[width=2.5in]{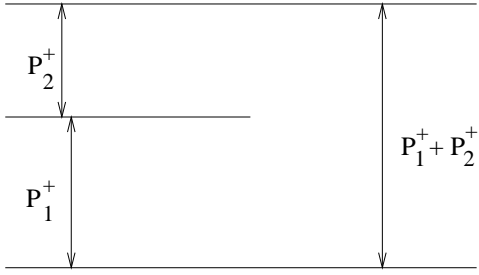}
\end{center}
\caption{The lightcone diagram for the three string vertex.}
\label{3vertex}
\end{figure}

It is a remarkable fact that the normalization of
diagrams implied by this simple prescription, defined concretely
by introducing a rectangular grid in $\sigma,\tau$, correctly reproduces
all of the multistring tree amplitudes of the dual resonance model.
This means in particular that the continuum limit of the worldsheet
lattice, introduced by Giles and Thorn \cite{gilest} (GT), 
is Lorentz covariant (in the critical dimension.
The simplest process which reflects this is the three string vertex
described by Fig.~\ref{3vertex}.
Since lightcone diagrams are properly normalized probability amplitudes,
Lorentz covariance dictates the $P^+$ dependence
\bea
{\rm Vertex} \sim {1\over\sqrt{P^+_1P^+_2(P^+_1+P^+_2)}}
\eea
for the vertex involving spin 0 states. This factor must come from
the determinant factor arising from the Gaussian integral over
${\bfs x}$. So under $P^+_i\to \lambda P^+_i$, which
is just a scaling of the size of the diagram,
the above diagram should scale as $\lambda^{-3/2}$.
A lightcone worldsheet lattice calculation gives $\lambda^{-(D-2)/16}$,
which makes clear the need for the critical dimension $D=26$ to
obtain Lorentz covariance \cite{gilest}. 

More generally the complete evaluation of a lightcone interacting
string diagram proceeds in two steps. First the dependence
on the initial and final data is 
extracted by shifting the ${\bfs x}(\sigma,\tau)$ 
by a solution of the classical equations of motion. The Gaussian integral
that remains is then expressed in terms of the determinant
of the Laplacian defined on the lightcone worldsheet.  
In this article we bring together in one place results 
on worldsheet determinants
scattered throughout the literature with the addition of some
new results and applications.

Central to our discussion will be an insightful
paper by Mark Kac studying what he called
``{Hearing the Shape of a  Drum}'' \cite{kacdrum}. His idea was
to connect the distribution of allowed normal mode frequencies $\lambda_k$,
which are the eigenvalues of the Laplacian $-\nabla^2/2$, to
the shape of a two dimensional membrane. Technically,
he considered a general polygonal shape and demonstrated
the small $t$ behavior of
\bea
\Tr\ e^{t\nabla^2/2}&=&\sum_k e^{-\lambda_k t}
\sim {{\rm Area}\over2\pi t}\mp{{\rm  Perimeter}\over
4\sqrt{2\pi t}}+\sum_{\rm corners}{1\over24}\left({\pi\over\theta_i}
-{\theta_i\over\pi}\right)+o(1)
\eea
where the minus sign is valid for Dirichlet and the plus sign for
Neumann boundary conditions on all edges.
Kac derived this formula for Dirichlet boundary conditions on all
polygon edges, but it also applies to Neumann boundary conditions on
all edges. If some edges have Dirichlet and others Neumann boundary
conditions, the perimeter is replaced by $L_D-L_N$.
The contribution of each $DD$ and $NN$ corner is as
above but $DN$ and $ND$ corners are different. The suitable 
generalization is given, for example in \cite{dowker}:
\bea
\Tr\ e^{t\nabla^2/2}\sim {{\rm Area}\over2\pi t}-{L_D-L_N\over
4\sqrt{2\pi t}}+\sum_{\rm {NN,DD\atop
corners}}{1\over24}\left({\pi\over\theta_i}
-{\theta_i\over\pi}\right)-\sum_{\rm {DN,ND\atop
corners}}{1\over48}\left({\pi\over\theta_i}
+{2\theta_i\over\pi}\right)+o(1).
\eea
In appendix A we confirm the $ND$ contribution 
in an elementary way for the special
cases $\theta=\pi/2M$, $M=1,2,3\ldots$, for which the method of images
can be successfully applied to the solutions of the diffusion equation.
Later on we show that
the general $\theta$ case is also a consequence of the $90^\circ$ case
(which can be inferred from the exact calculations for a rectangle
in the next section and the conformal transformation formula inferred from 
\cite{mckeans} and discussed in Section 3).

For the case of $90^\circ$ corners a $DD,NN$ corner contributes $+1/16$
whereas a $DN$ corner contributes $-1/16$.  
The lightcone open string vertex  
is a $360^\circ$ NN  corner. Putting $\theta=2\pi$,
\bea
{1\over24}\left({\pi\over\theta}-{\theta\over\pi}\right)\to-{1\over16}.
\eea
For 24 transverse dimensions the scaling power is thus $24/16=3/2$,
explaining the scaling law required by Lorentz covariance.
By taking an $n$ sided polygon with angles $\theta_i=\pi-\epsilon_i$ 
and the limit $n\to\infty$ with $\sum_i\epsilon_i=2\pi$, Kac showed that
a smooth closed curve will contribute a term to the above expression
of $1/6$. In particular a semi-circular arc subtending an angle $\theta$
would contribute a term $\theta/12\pi$.

In Section 2 we quote or derive expressions for the Gaussian path
integrals defined on a rectangular lattice. Most of these
results are known: see for example \cite{gilest}. We obtain
the results with all possible choices of Dirichlet (D) and
Neumann (N) boundary conditions. Each determinant can be expressed as
a single infinite product corresponding to diagonalizing
the transfer matrix in either the horizontal or vertical directions.
The equality of the two representations is a lattice analog of the
Jacobi imaginary transformation in the theory of elliptic functions.

In Section 3 we discuss some 
applications of the McKean-Singer result for the relation
of the determinants of the Laplacian on two regions related
by a conformal transformation. In Section 4 we review
Mandelstam's evaluation of the determinant for bosonic tree
diagrams and then discuss some possible interpretations
for the case of subcritical dimensions $D<26$, when some aspect
of Lorentz invariance fails. Also in Section 4 we discuss two applications
for string scattering. Technical details are relegated to two appendices.
\section{Rectangles: Lattice Results}
A brute force way to calculate worldsheet determinants is to
explicitly evaluate Gaussian path integrals on a lattice \cite{gilest}. 
So take an $M\times N$ finite rectangular lattice, with a coordinate
$x$ at each point on the lattice. Then we have
\bea
{\det}^{-1/2}(-\nabla^2)\to \int dx_{kl}\exp\left\{-{1\over2}\sum_{kl}
\left[(x_{k,l+1}-x_{k,l})^2+(x_{k+1,l}-x_{k,l})^2\right]\right\}
\label{gaussdef}
\eea
In each case $N,M,K$ are the number of integration variables in
a row of column of the lattice.

Points on the boundary of the
lattice can be fixed (Dirichlet) or freely integrated (Neumann).
The bilinear forms can be diagonalized by expanding in
normal modes, for which the eigenvalues are:
\bea
\alpha_n&\equiv&4\sin^2{n\pi\over2(N+1)},\qquad n=1,2,\ldots,N\\
\beta_n&\equiv&4\sin^2{m\pi\over2M},\qquad m=0,1,\ldots,M-1\\
\gamma_k&\equiv&4\sin^2{(k+1/2)\pi\over2K+1},\qquad k=0,1,\ldots,K-1
\eea
The $\alpha$'s are appropriate to a bilinear form with fixed ends (DD),
the $\beta$'s to a form with free ends (NN), and the $\gamma$'s
to a form with one fixed and one free end.
Then we are interested in the following determinants:
\bea
{\det}^{-1/2}_{\rm DDDD}&=&\prod_{n=1}^N\prod_{m=1}^{M}(\alpha_n+\alpha_m)^{-1/2},
\qquad{\det}^{-1/2}_{\rm DNDN}=
\prod_{n=1}^N\prod_{m=0}^{M-1}(\alpha_n+\beta_m)^{-1/2}\\
{\det}^{-1/2}_{\rm DDDN}&=&\prod_{n=1}^{N}
\prod_{k=0}^{K-1}(\alpha_n+\gamma_k)^{-1/2},\qquad
{\det}^{-1/2}_{\rm DNNN}=\prod_{m=0}^{M-1}
\prod_{k=0}^{K-1}(\beta_m+\gamma_k)^{-1/2}\\
{\det}^{-1/2}_{\rm DDNN}&=&\prod_{j=0}^{J-1}
\prod_{k=0}^{K-1}(\gamma_j+\gamma_k)^{-1/2}
\eea
In each of these formulas one of the products can be evaluated
exactly on the lattice.
The following product identities can be easily derived:
\bea
\prod_{n=1}^N(\alpha_n-z)={\sin(N+1)\kappa\over\sin\kappa},
\qquad\prod_{k=0}^{K-1}(\gamma_k-z)={\cos[(2K+1)\kappa/2]
\over\cos[\kappa/2]}
\eea
where $z$ and $\kappa$ are related by $z=4\sin^2[\kappa/2]$.
Applying these identities at $z=0,\kappa=0$ shows immediately
that $D_{\rm DNDN}=D_{\rm DDDD}/\sqrt{N+1}$ and $D_{\rm DNNN}=D_{\rm DDDN}$.
\subsection{DDDD}
We then find
\bea
{\det}^{-1/2}_{\rm DDDD}&=&\prod_{m=1}^M
\left[{\sinh(2(N+1)\sinh^{-1}(\sin(m\pi/2(M+1))))
\over\sinh(2\sinh^{-1}(\sin(m\pi/2(M+1))))}\right]^{-1/2}\nonumber\\
&=&(M+1)^{1/4}\left(
{\sinh[2(M+1)\sinh^{-1}1]\over2\sqrt{2}}\right)^{1/4}\nonumber\\
&&e^{-(N+1)\sum_{m=1}^M\sinh^{-1}\sin{m\pi/2(M+1)}}
\prod_{m=1}^M\left\{1-e^{-4(N+1)\sinh^{-1}\sin{m\pi/2(M+1)}}
\right\}^{-1/2}
\eea
where we used
\bea
\prod_{m=1}^M\left(2\sin{m\pi\over2(M+1)}\right)
&=&\sqrt{M+1}\\
\prod_{m=1}^M
\sqrt{4+4\sin^2{m\pi\over2(M+1)}}
&=&\left({\sinh(2(M+1)\sinh^{-1}1\over
\sinh(2\sinh^{-1}1)}\right)^{1/2}
\eea
and $\sinh(2\sinh^{-1}1)=2\sqrt{2}$.

The continuum limit is $M,N\to\infty$ with $L=(M+1)a$, and
$T=(N+1)a$ fixed. For this we need
\bea
\sum_{m=1}^M\sinh^{-1}\sin{m\pi\over2(M+1)}\sim 
{2(M+1)G\over\pi}-{1\over2}\sinh^{-1}1-{\pi\over24(M+1)}
\eea
where $G=\sum_{k=0}^\infty(-)^k/(2k+1)^2$ is Catalan's constant.
Then
\bea
{\det}^{-1/2}_{\rm DDDD}&\sim&\left({L\over 2a\sqrt{2}}\right)^{1/4}
e^{-\alpha LT + \beta(T+L)+\pi T/24L}
\prod_{m=1}^\infty\left\{1-e^{-2m\pi T/L}
\right\}^{-1/2}
\eea
with $\alpha\equiv 2G/\pi a^2$ and $\beta=(2a)^{-1}\sinh^{-1}1$.

We see that, apart from the coefficient $a^{-1/4}$, 
the divergences associated with the continuum limit
reside in the terms in the exponent 
proportional to the area or perimeter of
the rectangle. These terms are inconsequential and can be dropped in order
to define a finite continuum determinant
\bea
{\det}^{-1/2}_{DDDD,C}&\equiv&{L}^{1/4}e^{\pi T/24L}
\prod_{m=1}^\infty\left\{1-e^{-2m\pi T/L}
\right\}^{-1/2}
\eea
The factor of $L^{1/4}$ accounts for the corner contribution in the
Kac formula, in this case 4 $90^\circ$ corners or $4\times(1/16)$.
The remaining factors depend on the shape $T/L$ of the rectangle.
The symmetry $T\leftrightarrow L$ of the rectangle and boundary conditions 
implies the equality
\bea
{L}^{1/4}e^{\pi T/24L}
\prod_{m=1}^\infty\left\{1-e^{-2m\pi T/L}
\right\}^{-1/2}&=&{T}^{1/4}e^{\pi L/24T}
\prod_{m=1}^\infty\left\{1-e^{-2m\pi L/T}
\right\}^{-1/2}
\eea
which is simply the Jacobi transform in the theory of elliptic 
functions.\footnote{
In standard notation with $q\equiv e^{i\pi\tau}=e^{-2\pi T/L}$
and $\dot q=e^{i\pi\dot\tau}$ this identity reads
\bea
q^{-1/48}\prod_{n=1}^\infty(1-q^n)^{-1/2}
&=&\left({-i\tau\over2\pi}\right)^{1/4}\dot{q}^{-1/48}
\prod_{n=1}^\infty(1-\dot{q}^n)^{-1/2}.
\eea}
\subsection{DNDN}
From the identity 
${\det}^{-1/2}_{\rm DNDN}={\det}^{-1/2}_{\rm DDDD}/\sqrt{N+1}$
we can immediately write down
\bea
{\det}^{-1/2}_{DNDN,C}&\equiv&{L}^{-1/4}\sqrt{L/T}e^{\pi T/24L}
\prod_{m=1}^\infty\left\{1-e^{-2m\pi T/L}\right\}^{-1/2}
\eea
The scaling power is now $-1/4$ corresponding to 4 $90^\circ$ ND
corners in the Kac formula. Because the boundary conditions break
the symmetry $T\leftrightarrow L$ the determinant doesn't 
have the symmetry:
\bea
{L}^{-1/4}\sqrt{L/T}e^{\pi T/24L}
\prod_{m=1}^\infty\left\{1-e^{-2m\pi T/L}\right\}^{-1/2}
={T}^{-1/4}e^{\pi L/24T}
\prod_{m=1}^\infty\left\{1-e^{-2m\pi L/T}\right\}^{-1/2}
\eea
The factor $\sqrt{L/T}$ on the left reflects the propagation in
$T$ of the zero mode of an NN string. The right shows the propagation
in $L$ which is that of a DD string with no zero mode.
\subsection{DDDN and DNNN}
Next we turn to the $DDDN$ determinant. Doing the product over $n$,
we find
\bea
{\det}^{-1/2}_{\rm DDDN}&=&\prod_{k=0}^{K-1}
\left[{\sinh(2(N+1)\sinh^{-1}(\sin((k+1/2)\pi/(2K+1))))
\over\sinh(2\sinh^{-1}(\sin((k+1/2)\pi/(2K+1))))}\right]^{-1/2}\nonumber\\
&=&\left(
{\cosh[(2K+1)\sinh^{-1}1]\over\sqrt{2}}\right)^{1/4}
e^{-(N+1)\sum_{k=0}^{K-1}\sinh^{-1}\sin{(k+1/2)\pi/(2K+1)}}\nonumber\\
&&
\prod_{k=0}^{K-1}\left\{1-e^{-4(N+1)\sinh^{-1}\sin{(k+1/2)\pi/(2K+1)}}
\right\}^{-1/2}
\eea
where we used
\bea
\prod_{k=0}^{K-1}\left(2\sin{(k+1/2)\pi\over2K+1}\right)
&=&1\\
\prod_{k=0}^{K-1}
\sqrt{4+4\sin^2{(k+1/2)\pi\over2K+1}}
&=&\left({\cosh((2K+1)\sinh^{-1}1\over
\cosh(\sinh^{-1}1)}\right)^{1/2}
\eea
and $\cosh(\sinh^{-1}1)=\sqrt{2}$. For the continuum limit we need
\bea
\sum_{k=0}^{K-1}\sinh^{-1}\sin{(k+1/2)\pi\over2K+1}\sim 
{(2K+1)G\over\pi}-{1\over2}\sinh^{-1}1+{\pi\over48(K+1/2)}
\eea
With the understanding that the ''length'' of an ND string is $L=a(K+1/2)$
we see the bulk and boundary terms are identical to the DDDD and
DNDN cases. So the continuum limit is
\bea
{\det}^{-1/2}_{\rm DDDN}&=&
{2^{-3/8}}e^{-\alpha LT+\beta(L+T)-\pi T/48L}
\prod_{k=0}^\infty\left\{1-e^{-2(k+1/2)\pi T/L}
\right\}^{-1/2}
\eea
with $\alpha=2G/\pi a^2$ and $\beta=(2a)^{-1}\sinh^{-1}1$ as before.
The corresponding continuum determinant can be taken to be
\bea
{\det}^{-1/2}_{\rm DDDN,C}&=&
e^{-\pi T/48L}
\prod_{k=0}^\infty\left\{1-e^{-2(k+1/2)\pi T/L}
\right\}^{-1/2}
\eea
This determinant is scale invariant in accord with the fact that this
rectangle has two ND $90^\circ$ corners and 2 DD $90^\circ$ corners
which contribute with cancelling signs. In this form the determinant
displays the propagation of a DN string in $T$. A Jacobi transform
displays the propagation of a DD string in $L$\footnote{In terms of
$q,\dot q$ defined in the previous footnote this identity reads
\bea
q^{1/96}\prod_{k=0}^\infty(1-q^{k+1/2})^{-1/2}
&=&2^{-1/4}\dot{q}^{-1/48}
\prod_{n=1}^\infty(1+\dot{q}^n)^{-1/2}.
\eea}:
\bea
e^{-\pi T/48L}
\prod_{k=0}^\infty\left\{1-e^{-2(k+1/2)\pi T/L}\right\}^{-1/2}
&=&2^{-1/4}e^{\pi L/24T}\prod_{n=1}^\infty(1+e^{-2n\pi L/T})^{-1/2}
\eea
We have already noted that the determinant for the NNND case is identical
to the DDDN case we just discussed.
\subsection{DDNN}
Finally for completeness we analyze the DDNN rectangle, which
reflects a DN string propagating in both $T$ and $L$. Like the
DDDD case the result should possess the symmetry $T\leftrightarrow L$.
Doing the product over $j$ gives
\bea
{\det}^{-1/2}_{\rm DDNN}&=&\prod_{k=0}^{K-1}
\left[{\cosh((2J+1)\sinh^{-1}(\sin((k+1/2)\pi/(2K+1))))
\over\cosh(\sinh^{-1}(\sin((k+1/2)\pi/(2K+1))))}\right]^{-1/2}\nonumber\\
&=&\left(
{\cosh[(2K+1)\sinh^{-1}1]\over\sqrt{2}}\right)^{1/4}
e^{-(J+1/2)\sum_{k=0}^{K-1}\sinh^{-1}\sin{(k+1/2)\pi/(2K+1)}}\nonumber\\
&&
\prod_{k=0}^{K-1}\left\{1+e^{-2(2J+1)\sinh^{-1}\sin{(k+1/2)\pi/(2K+1)}}
\right\}^{-1/2}\\
&\sim&{2^{-3/8}}e^{-\alpha LT+\beta(L+T)-\pi T/48L}
\prod_{k=0}^\infty\left\{1+e^{-2(k+1/2)\pi T/L}
\right\}^{-1/2}
\eea
where the last line is the continuum limit with the identifications
$L=K+1/2$ and $T=J+1/2$. Just as in the previous cases the continuum
determinant can then be chosen as
\bea
{\det}^{-1/2}_{DDNN,C}&=&e^{-\pi T/48L}
\prod_{k=0}^\infty\left\{1+e^{-2(k+1/2)\pi T/L}
\right\}^{-1/2}
\eea
And the symmetry under $T\leftrightarrow L$ is yet another Jacobi
relation
\bea
e^{-\pi T/48L}
\prod_{k=0}^\infty\left\{1+e^{-2(k+1/2)\pi T/L}
\right\}^{-1/2}&=&e^{-\pi L/48T}
\prod_{k=0}^\infty\left\{1+e^{-2(k+1/2)\pi L/T}
\right\}^{-1/2}
\eea
A noteworthy feature of the lattice definition of the various
determinants is that the bulk and boundary terms are identical in
all cases: $-\alpha LT +\beta(L+T)$ regardless of how Dirichlet
and Neumann conditions are assigned. This contrasts with the
diffusion equation method of Kac and McKean-Singer. Of course
this statement entails a varying identification between the
continuum lengths and the number of degrees of freedom:
$L/a=M,N+1,K+1/2$ for NN, DD, DN conditions respectively,
and similarly for $T$. That is, there is an intrinsic ambiguity in
identifying a unique ``continuum'' length. If we express these three
lengths in terms of the ND one $L_0=a(K+1/2)$, they are
$L_0-a/2$, $L_0$, and $L_0+a/2$. This is a variation that mirrors
the results of the diffusion equation continuum method.
\subsection{NNNN}
We end this section with a brief aside on the NNNN case, which requires
special handling because of the zero mode. This zero mode is due to the
translational invariance of the Gaussian integral 
(\ref{gaussdef}) that we have used to define
determinants. To interpret it add a source term $i\sum_{kl}x_{kl}J_{kl}$
to the exponent. Then insert (a la Fadeev-Popov)
\bea
1=\int da\delta\left(a-{1\over MN}\sum_{kl}x_{kl}\right)
\eea
in the $x$ integrand. A change of variables $x_{kl}\to x_{kl}+a$ transfers 
the $a$ dependence to the exponent and then integration over $a$
produces a delta function factor
\bea
\int da e^{ia\sum_{kl}J_{kl}}=2\pi\delta(\sum_{kl}J_{kl}).
\label{conserve}\eea
Since $J$ is conjugate to $x$, we see that this factor simply enforces
momentum conservation. This interprets the infinite factor due
to translation invariance as $2\pi\delta(0)=\infty$. The coefficient 
of the delta function has a finite zero source limit $J_{kl}\to0$.
We define $\det^{-1/2}_{\rm NNNN}$ as this coefficient at zero source:
\bea
{\det}^{-1/2}_{\rm NNNN}\equiv\int dx_{kl}
\delta\left({1\over MN}\sum_{kl}x_{kl}\right)\exp\left\{-{1\over2}\sum_{kl}
\left[(x_{k,l+1}-x_{k,l})^2+(x_{k+1,l}-x_{k,l})^2\right]\right\}
\eea
When we change variables to normal modes $q_{mn}$, normalized so
that  the Jacobian is unity, we find that $q_{00}\sqrt{MN}=\sum_{kl} x_{kl}$.
Hence the effect of the delta function is to multiply by $\sqrt{MN}$
and delete the contribution of the zero mode:
\bea
{\det}^{-1/2}_{\rm NNNN}&=&\sqrt{MN}
\prod_{(m,n)\neq(0,0)}(\beta_m+\beta_n)^{-1/2}\nonumber\\
&=&\sqrt{MN}\prod_{m=1}^{M-1}\beta_m^{-1/2}\prod_{n=1}^{N-1}\beta_n^{-1/2}
{\det}^{-1/2}_{\rm DDDD}={\det}^{-1/2}_{\rm DDDD}
\eea
This formula confirms that NN $90^\circ$ corners have identical effect to DD
$90^\circ$ corners.
\section{Conformal Transformation}
More generally, under a conformal scaling $g_{ab}\to e^{2\Sigma}g_{ab}$,
the change in the determinant of the Laplacian is given by \cite{mckeans}
\bea
-{1\over2}\delta(\ln(-\nabla^2))&=&{1\over 24\pi}\int dA g^{ab}
{d\Sigma\over dz_a}{d\Sigma\over dz_b}+{1\over12\pi}\int d\ell k\Sigma
+{1\over24\pi}\int dA R\Sigma\nonumber\\
&&+{1\over24}\sum_{\rm {DD, NN\atop corners}}\left({\pi\over\theta_i}-
{\theta_i\over\pi}\right)\Sigma(z_i)-{1\over48}\sum_{\rm {DN\atop corners}}
\left({\pi\over\theta_i}+{2\theta_i\over\pi}\right)\Sigma(z_i)
\label{conformal}
\eea
Here it is understood that the two determinants have the bulk and
boundary terms dropped. Actually this formula does not
explicitly appear in \cite{mckeans}. Rather the first three terms
in the asymptotic behavior as $t\to0$ of 
$\Tr e^{t\nabla^2}$ are explicitly calculated in terms of the
geometry of an
arbitrary smooth manifold endowed with a metric $g_{ab}$. The change
formula then follows after a straightforward evaluation  of
the difference of their results for two manifolds
related by a conformal transformation 
(see, for example \cite{alvarez}). When the boundary
is only piece-wise smooth, the corner terms that appear can be  
inferred from Kac's results, and their generalization to DN
corners.

\subsection{DD Corners from Conformal Transform of a Rectangle}
\begin{figure}[ht]
\begin{center}
\includegraphics[width=5in]{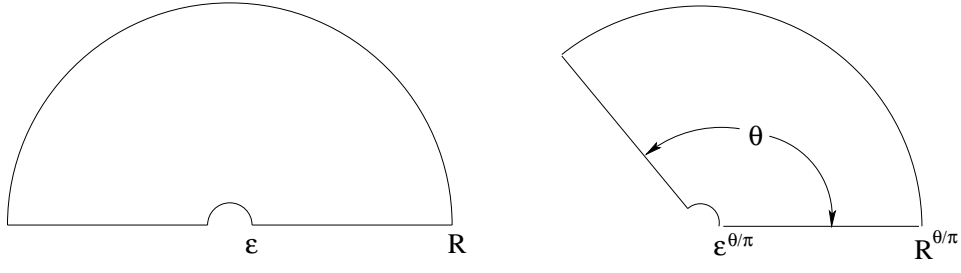}
\caption{Two geometries related by the conformal transformation
$y=z^{\theta/\pi}$.}
\label{annuli}
\end{center}
\end{figure}

We can use (\ref{conformal}) to obtain the measure for the region on the
right of Fig.~\ref{annuli} from the measure for the figure on the
left, or, more fundamentally, from the measure for a rectangle,
which we have explicitly evaluated in Section 2 by introducing a rectangular
lattice.
The shapes in Fig.~\ref{annuli} are related by the transformation
\bea
y&=&{\pi\over\theta}z^{\theta/\pi},\qquad 
\Sigma=\ln\left|{dy\over dz}\right|=\left({\theta\over\pi}-1\right)\ln|z|
\eea
However, we begin by recalling a formula
for the determinant of an $M\times N$ rectangular
grid \cite{gilest}.
\bea
{\det}^{-1/2}(-\nabla^2)&=&(M+1)^{1/4}\left(
{\sinh[2(M+1)\sinh^{-1}1]\over\sinh[2\sinh^{-1}1]}\right)^{1/4}e^{-(N+1)\sum_{m=1}^M\sinh^{-1}\sin{m\pi/2(M+1)}}\nonumber\\
&&\times
\prod_{m=1}^M\left\{1-e^{-4(N+1)\sinh^{-1}\sin{m\pi/2(M+1)}}
\right\}^{-1/2}
\eea
In the continuum limit $M,N\to\infty$, with $T/L\equiv(N+1)/(M+1)$
fixed, this reduces to:
\bea
{\det}^{-1/2}(-\nabla^2)&\sim&Ke^{\alpha LT+\beta(L+T)}L^{1/4}
e^{\pi T/24L}\prod_{m=1}^\infty\left\{1-e^{-2m\pi T/L}\right\}^{-1/2}
\eea
The factor of $L^{1/4}$ reflects the scaling predicted by Kac
for the four $90^\circ$ corners of the rectangle. We may therefore choose
a standard rectangle setting $L=\pi$ and 
dropping the divergent area and perimeter terms in the exponential prefactor, 
\bea
{\det}^{-1/2}_{\rm DDDD~rect}(-\nabla^2)&\sim&
e^{T/24}\prod_{m=1}^\infty\left\{1-e^{-2mT}\right\}^{-1/2}
\eea
It is convenient to coordinatize the rectangle by the
complex variable $\rho=\tau+i\sigma$, with $0<\sigma<\pi$ and
$T_1<\tau<T_2$. Then the conformal transformation,
$z=e^{\theta\rho\over\pi}$ maps the rectangle onto a wedge of
an annulus of angle $\theta$ inner radius $\epsilon=e^{\theta T_1/\pi}$
and outer radius $R=e^{\theta T_2/\pi}$. To get the determinant for this
new region, we first compute
\bea
{dz\over d\rho}&=&{\theta\over\pi}e^{\theta\rho\over\pi},\qquad
\Sigma=\ln{\theta\over\pi}+{\theta\over\pi}{\rm Re}\ \rho,\qquad
\partial_n\Sigma=\cases{\theta/\pi & $\tau=T_2$\cr
-\theta/\pi& $\tau=T_1$\cr 0 &$\sigma=0,\pi$\cr}\\
{1\over24\pi}\oint d\rho \Sigma\partial_n\Sigma&=&
{\theta^2\over24\pi^2}(T_2-T_1)
={\theta\over24\pi}\ln{R\over\epsilon}
\eea
From the $90^\circ$ corners we need
\bea
{1\over 16}\sum_i\Sigma_i={1\over8}\left({\theta\over\pi}(T_1+T_2)
+2\ln{\theta\over\pi}\right)={1\over8}\left(\ln R + \ln\epsilon
+2\ln{\theta\over\pi}\right)
\eea
Thus we have 
\bea
-{1\over2}\ln{\det}_{\rm DD~annular~wedge}
&=&-{1\over2}\ln{\det}_{\rm DDDD~rect}+{\theta\over24\pi}\ln{R\over\epsilon}
+{1\over8}\left(\ln R + \ln\epsilon
+2\ln{\theta\over\pi}\right)\nonumber\\
&&\hskip-.75in
={T_2-T_1\over24}-{1\over2}\sum_{m=1}^\infty\ln(1-e^{-2m(T_2-T_1)})
+{\theta\over24\pi}\ln{R\over\epsilon}
+{1\over8}\left(\ln {\theta R\over\pi} + \ln{\theta\epsilon\over\pi}
\right)\nonumber\\
&&\hskip-.75in
=-{1\over2}\sum_{m=1}^\infty\ln(1-(\epsilon/R)^{2m\pi/\theta})
+{1\over24}\left({\pi\over\theta}+{\theta\over\pi}\right)\ln{R\over\epsilon}
+{1\over8}\left(\ln {\theta R\over\pi} + \ln{\theta\epsilon\over\pi}
\right)\nonumber\\
&&\hskip-.75in
\sim{1\over24}\left({\pi\over\theta}+{\theta\over\pi}\right)\ln{R\over\epsilon}
+{1\over8}\left(\ln {\theta R\over\pi} + \ln{\theta\epsilon\over\pi}
\right)
\eea
where the last line is valid for $\epsilon\ll R$. If we drop the 
$\epsilon$ terms in this limit, the $R$ terms that remain should
give the determinant for the wedge with the annular hole removed
\bea
-{1\over2}\ln{\det}_{\rm DD~wedge}&\sim&{1\over24}
\left({\pi\over\theta}+{\theta\over\pi}\right)\ln{R}
+{1\over8}\ln R\nonumber\\
&=&{1\over24}
\left({\pi\over\theta}-{\theta\over\pi}+3\right)\ln{R}
+{\theta\over12\pi}\ln R
\eea
where we also dropped the scale independent term $(1/4)\ln(\theta/\pi)$.
The first term agrees with Kac's formula for corners: one corner
of angle $\theta$ and two corners of angle $\pi/2$. The last term
is the contribution from the circular arc, which we have seen 
follows from the Kac formula for a limiting polygon with corner
angles $\sim \pi$. In this way we see that the conformal transformation
formula embodies Kac's result as well as it's McKean-Singer generalization. 
\subsection{DN Corners from Conformal Transform of a Rectangle.}
If we replace the DDDD rectangle used in the previous subsection
with a DDDN rectangle we learn about DN corners. In that case
the corner contributions to the conformal transformation formula
cancel and we have 
\bea
&&\hskip-2cm -{1\over2}\ln{\det}_{\rm DN~annular~wedge}\nonumber\\
&=&-{1\over2}\ln{\det}_{\rm DDDN~rect}+{\theta\over24\pi}\ln{R\over\epsilon}
\nonumber\\
&=&-{T_2-T_1\over48}-{1\over2}\sum_{k=0}^\infty\ln(1-e^{-(2k+1)(T_2-T_1)})
+{\theta\over24\pi}\ln{R\over\epsilon}
\nonumber\\
&=&-{1\over2}\sum_{k=0}^\infty\ln(1-(\epsilon/R)^{(2k+1)\pi/\theta})
+{1\over24}\left(-{\pi\over2\theta}+{\theta\over\pi}\right)\ln{R\over\epsilon}
\nonumber\\
&\sim&{1\over24}\left(-{\pi\over2\theta}+{\theta\over\pi}\right)
\ln{R\over\epsilon},\qquad {\epsilon\over R} \to 0.
\eea
Dropping the $\epsilon$ terms produces the determinant for a
wedge of angle $\theta$
\bea
 -{1\over2}\ln{\det}_{\rm DN~wedge}&=&{1\over24}\left(-{\pi\over2\theta}+{\theta\over\pi}\right)
\ln{R}=-{1\over24}\left({\pi\over2\theta}+{\theta\over\pi}\right)\ln{R}
+{\theta\over12\pi}\ln R.
\eea
Again the last term accounts for the contribution from the circular
arc that closes the wedge, whence the first term must be associated
with the DN angle itself. (The corners at the end of the
arc contribute opposite signs and cancel.)
It is seen to agree with our generalized Kac formula. 
\section{Lightcone Bosonic Tree}
From Mandelstam's work \cite{mandelstamdet}, 
the measure factor for an $N$ point tree is
\bea
\left|{\partial T\over\partial Z}\right| {\det}^{-(D-2)/2}(-\nabla^2)
&=& Z_{N-1}\prod_{k=1}^N{1\over |
\alpha_k|^{(D-2)/48}}\left[|\alpha_N|^{N-3}{\prod_{r<t}|x_t-x_r|
\over \prod_{m<l}|Z_l-Z_m|}\right]^{(26-D)/24}
\eea
where $\alpha_r=2p^+_r$.
The rest of the Koba-Nielsen integrand is just the usual
$\prod_{i<j}|Z_j-Z_i|^{2k_i\cdot k_j}$ (in units where $\alpha^\prime=1$)
and for which $k_j^2=(D-2)/24$.

The quantities $Z_k$, with $k=1\cdots (N-1)$, and $x_r$, 
with $r=1\cdots(N-2)$ are determined from the map from the
upper-half Koba-Nielsen plane ($z$) to the lightcone world sheet
($\rho=\tau+i\sigma$):
\bea
\rho&=&\sum_{k=1}^{N-1}\alpha_k\ln(z-Z_k),\qquad {d\rho\over dz}\bigg|_{z=x_r}=0\\
{d\rho\over dz}&=&\sum_{k=1}^{N-1}{\alpha_k\over z-Z_k}
={\sum_{k=1}^{N-1}\alpha_k\prod_{l\neq k}(z-Z_l)\over
\prod_k(z-Z_k)}=-\alpha_N{\prod_r (z-x_r)\over\prod_k(z-Z_k)}
\eea
so that the asymptotic strings at $\tau=\pm\infty$ are mapped from
the $Z_k$. In this notation $Z_N=\infty, Z_1=0$. 
A useful identity follows by setting $z=Z_m$ in the identity
\bea
-\alpha_N\prod_r (z-x_r)&=&\sum_{k=1}^{N-1}\alpha_k\prod_{l\neq k}(z-Z_l)\\
-\alpha_N\prod_r (Z_m-x_r)&=&\sum_{k=1}^{N-1}\alpha_k\prod_{l\neq k}(Z_m-Z_l)
=\alpha_m\prod_{l\neq m}(Z_m-Z_l)\\
|\alpha_N|^{N}\prod_{m,r}|Z_m-x_r|&=&\prod_{m=1}^{N}|\alpha_m|
\prod_{l\neq k}|Z_k-Z_l|
\eea
Then the measure can be put in the more suggestive form
\bea
&&\hskip-1.5cm\left|{\partial T\over\partial Z}\right| 
{\det}^{-(D-2)/2}(-\nabla^2)\nonumber\\
&=& Z_{N-1}\prod_{k=1}^N{1\over |
\alpha_k|^{(D-2)/48}}\left[|\alpha_N|^{N-3}{\prod_{r<t}|x_t-x_r|
\prod_{m<l}|Z_l-Z_m|
\over \prod_{m\neq l}|Z_l-Z_m|}\right]^{(26-D)/24}\nonumber\\
&=&Z_{N-1}\prod_{k=1}^N{1\over |
\alpha_k|^{(D-2)/48}}\left[{\prod_k|\alpha_k|\over
|\alpha_N|^{3}}{\prod_{r<t}|x_t-x_r|
\prod_{m<l}|Z_l-Z_m|
\over \prod_{l,r}|Z_l-x_r|}\right]^{(26-D)/24}\nonumber\\
&=&Z_{N-1}\prod_{k=1}^N{1\over \sqrt{|
\alpha_k|}}\left[{\prod_k|\alpha_k|^{3/2}\over
|\alpha_N|^{3}}{\prod_{r<t}|x_t-x_r|
\prod_{m<l}|Z_l-Z_m|
\over \prod_{l,r}|Z_l-x_r|}\right]^{(26-D)/24}\\
&=&Z_{N-1}\prod_{k=1}^N{1\over \sqrt{|
\alpha_k|}}\left[{\prod_{k<N}|\alpha_k|\over
|\alpha_N|}\right]^{(26-D)/16}\left[{\prod_{r<t}|x_t-x_r|
\prod_{m<l}|Z_l-Z_m|
\over \prod_{l,r}|Z_l-x_r|}\right]^{(26-D)/24}
\eea
\subsection{Interpretation of $D<26$}
The factors in square brackets spoil Lorentz covariance for
$D<26$. However the $x_r, Z_k$ dependence of these factors
is in a form that can be cancelled by inserting an operator
of the form $e^{\pm i\gamma \phi(\rho)}$ at each $x_r, Z_k$. 
Here $\phi(\rho)$ is one of the transverse string coordinates.
Take the $D$ indices of $x^\mu$ to be $0,1,2,\cdots,(D-1)$.
Then $x^\pm=(x^0\pm x^1)/\sqrt2$ and the transverse components
are $2,\cdots (D-1)$. We choose $\phi(\rho)=x^{D-1}$.
Clearly inserting such operators sacrifices the full $SO(D-1,1)$
Lorentz invariance. But if the Lorentz violating measure can indeed
be cancelled in this way, the scattering amplitudes will 
be invariant under $SO(D-2,1)$ Lorentz invariance. For example,
to get a subcritical string theory that respects $3+1$ Lorentz invariance,
we should start with $5=4+1$ dimensional space-time.

The contribution of the field $\phi$ to the Boltzmann factor
of the worldsheet path integral
will be
\bea
B(\phi)&=&\exp\left\{-{1\over4\pi}\int d^2\rho (\nabla\phi)^2
+i\gamma\sum_r \phi(\rho(x_r))
+i\sum_k {p_k\over2\pi p_k^+}\int d\sigma_k \phi(\sigma_k,\tau_k)\right\}
\eea
The last term converts the initial and final state description from
coordinate space to momentum space, and we specialize to a constant
momentum density on each string at initial and final times.
Here, to conform with Mandelstam's (and also GT's) conventions, 
we have taken $\alpha^\prime=1$ and scale the worldsheet spatial
coordinate $\sigma_{old}=T_0\sigma_{new}=\sigma_{new}/2\pi$ so that on a
given string $0<\sigma_{new}<2\pi p_k^+\equiv \pi\alpha_k$.
Thus $\rho=\tau+i\sigma_{new}$, and henceforth $\sigma=\sigma_{new}$.

We can extract the $\gamma$ dependence of the path integral
by completing the square in the usual way. We shift
$\phi\to\phi+c$ and choose $c$ to cancel the linear terms:
\bea
-\nabla^2 c&=&2\pi i\gamma\sum_r\delta(\rho-\rho(x_r)),\qquad
{\dot c}|_{\tau_f}=i{p_k\over p^+_k},
\quad {\dot c}|_{\tau_i}=-i{p_k\over p^+_k},\qquad c^\prime|_{\partial}=0\\
\ln {B(\phi+c)\over B(\phi)|_{0}}&=&
+{i\gamma\over2}\sum_s c(x_s)+{i\over2}
\sum_k{p_k\over 2\pi p^+_k}\int d\sigma_k c(\sigma_k,\tau_k)
\eea
The answer can be expressed in terms of the Neumann function
\bea
-\nabla^2N(\rho,\rho^\prime)=-2\pi\delta(\rho-\rho^\prime),\qquad
\partial_n N(\rho,\rho^\prime)|_{\rho\in\partial}=f(\rho)
\eea
Then applying Green's theorem we have
\bea
c(\rho^\prime)&=&-i\gamma\sum_r N(\rho(x_r),\rho^\prime)
-i\sum_{k\in f}{p_k\over 2\pi p^+_k}\int d\sigma_kN(\rho,\rho^\prime)
+i\sum_{k\in i}{p_k\over 2\pi p^+_k}\int d\sigma_kN(\rho,\rho^\prime)
\nonumber\\
&&+{1\over2\pi}\int d\sigma (cf)\bigg|^{\tau_f}_{\tau_i}
\eea
The last term, independent of $\rho^\prime$ drops out of $\ln B/B_0$:
\bea
\ln {B(\phi+c)\over B(\phi)|_{0}}&=&{\gamma^2\over2}\sum_{r,s}
N(\rho(x_r),\rho(x_s))+\gamma\sum_{r,k}{p_k\over2\pi p^+_k}
\int d\sigma_k N(\rho(x_r),\rho_k)\nonumber\\
&&+{1\over2}\sum_{kl}
\int d\sigma_k d\sigma_l{p_k p_l\over4\pi^2p^+_kp^+_l}N(\rho_k,\rho_l)
\eea
The Neumann function on the upper half plane is
\bea
N(z,z^\prime)=\ln|z-z^\prime|+\ln|z-z^{\prime*}|\to 2\ln|z-z^\prime|
\eea
when one or both $z$'s are on the real axis. Then, with $z(\rho)$ the
conformal map from the string diagram to the upper half plane
we find
\bea
\ln {B(\phi+c)\over B(\phi)|_{0}}&=&{\gamma^2}\sum_{r\neq s}
\ln|x_r-x_s|+2\gamma\sum_{r,k}p_k\ln|x_r-Z_k|
+\sum_{k\neq l}p_k p_l\ln|Z_k-Z_l|\nonumber\\
&&+\gamma^2\sum_r\ln|x_r-x_r| +{1\over2}\sum_{k}
\int d\sigma_k d\sigma^\prime_k{p^2_k\over4\pi^2p^{+2}_k}
N(\rho_k,\rho^\prime_k)
\eea
Note that $Z_N$, which we have set to $\infty$, appears on the
right side in the combination
\bea
2p_N(\gamma+\sum_{k=1}^{N-1}p_k)\ln Z_N=-2p_N^2\ln Z_N
\eea
so the terms involving $Z_N$ for this special dimension will
combine just as with the other dimensions into the terms
that lead to the mass shell condition on the $N$th leg. We
therefore can drop them.
The self-contractions on the last line need further discussion.
Those in the last term are of the same form for all transverse dimensions
and combined give the mass shell condition. Let us denote the
first $D-3$ transverse components as a vector ${\bfs p}_k$ in bold face
type, retaining roman type for the last one. Then the lightcone
mass shell condition reads
\bea
{\bfs p}_k^2-2p^+_kp^-_k={D-2\over24}-p_k^2
\eea
The left side of this equation is Lorentz invariant $p_{k\mu}p^\mu_k$
which should be $+1$ to describe the subcritical Veneziano model.
This requires that $p_k^2=(D-26)/24$, or $p_k=\pm i\sqrt{(26-D)/24}$.
In this case the requirement
\bea
\sum_k p_k=-(N-2)\gamma
\eea
can be met if $p_k=-i\sqrt{(26-D)/24}=-\gamma$ for $N-1$ values of $k$
and the $N$th momentum is $+i\sqrt{(26-D)/24}=+\gamma$.

Finally we need an interpretation of the self contractions at
the interaction points. Infinities in these contractions can be
absorbed into the coupling constant, provided they are independent of the
geometry of the worldsheet. Since the lightcone worldsheet is
the fundamental starting point, we should set any regulator cutoffs 
in the $\rho$ coordinate. Let us examine $\rho(z)$ near $z=x_r$,
where $d\rho/dz=0$:
\bea
\rho(z)&\approx& \rho(x_r) +{1\over2}{d^2\rho\over dz^2}\bigg|_{z=x_r}
(z-x_r)^2\\
{d^2\rho\over dz^2}\bigg|_{z=x_r}&=&-\alpha_N{\prod_{s\neq r}(x_r-x_s)\over
\prod_{k}(x_r-z_k)}\\
z-x_r&\approx&\sqrt{\rho-\rho(x_r)}
\left[{\prod_{k}(x_r-z_k)\over-\alpha_N\prod_{s\neq r}(x_r-x_s)}
\right]^{1/2}\\
|z(\rho)-z(\rho^\prime)|&\approx&|\sqrt{\rho-\rho(x_r)}
-\sqrt{\rho^\prime-\rho(x_r)}|{\prod_{k}\sqrt{|x_r-z_k|}\over\sqrt{|\alpha_N|}
\prod_{s\neq r}\sqrt{|x_r-x_s|}}\\
\gamma^2\sum_e\ln|x_r-x_r|&\to&{\gamma^2\over2}\left[(N-2)\ln\epsilon+\ln{\prod_{r,k}
{|x_r-Z_k|}\over{|\alpha_N|^{N-2}}
\prod_{s\neq r}{|x_r-x_s|}}\right]\\
&&\hskip-2cm\to{\gamma^2\over2}\left[(N-2)\ln\epsilon+\ln{|\alpha_N|\over
\prod_{k=1}^{N-1}|\alpha_k|}+\ln{\prod_{r,k}
{|x_r-Z_k|^2}\over
\prod_{s\neq r}{|x_r-x_s|}\prod_{k\neq l}|Z_k-Z_l|}\right]
\eea
The last line is our interpretation of the self contractions at
the interaction points
where we have let $\epsilon$ be a measure of the cutoff regularization on the
lightcone world sheet.

Having taken care of the terms involving $Z_N$ and the self contractions
at the external states, and setting $p_k=-\gamma$, for $k<N$ and
$p_N=+\gamma$,
what is left of the contribution from the
insertion operator is the correction factor
\bea
C&=&\epsilon^{(N-2)\gamma^2/2}\left[{\prod_{r\neq s}|x_r-x_s|
\prod_{k\neq l<N}|Z_k-Z_l|
\over\prod_{r,k<N}|x_r-Z_k|^2}\right]^{\gamma^2/2}\left[
{|\alpha_N|\over\prod_{k=1}^{N-1}|\alpha_k|}\right]^{\gamma^2/2}\nonumber\\
&=&\epsilon^{(N-2)\gamma^2/2}\left[{\prod_{r<s}|x_s-x_r|
\prod_{l< k<N}|Z_k-Z_l|
\over\prod_{r,k<N}|x_r-Z_k|}\right]^{\gamma^2}\left[
{\prod_{k=1}^{N-1}|\alpha_k|\over|\alpha_N|}\right]^{-\gamma^2/2}
\eea
More generally we can choose another momentum $p_n=+\gamma$, with
$p_k=-\gamma$ for $k\neq n$. In that case the terms in $\ln C$
{\it linear} in $p_n$ change sign, that is
\bea
-2\gamma^2\sum_r\ln|x_r-Z_n|+2\gamma^2\sum_{k\neq n, N}\ln|Z_k-Z_n|
&&\nonumber\\
&&\hskip-2in\to
+2\gamma^2\sum_r\ln|x_r-Z_n|-2\gamma^2\sum_{k\neq n, N}\ln|Z_k-Z_n|\nonumber\\
&&\hskip-2in=
-2\gamma^2\sum_r\ln|x_r-Z_n|+2\gamma^2\sum_{k\neq n, N}\ln|Z_k-Z_n|
+4\gamma^2\ln{|\alpha_n|\over|\alpha_N|}
\eea
where we have used the identity $-\alpha_N\prod_r(Z_n-x_r)
=\alpha_n\prod_{k\neq n,N}(Z_n-Z_k)$, which we have proven earlier.
Thus in more generality the correction factor becomes
\bea
C&=&\epsilon^{(N-2)\gamma^2/2}\left[{\prod_{r<s}|x_s-x_r|
\prod_{l< k<N}|Z_k-Z_l|
\over\prod_{r,k<N}|x_r-Z_k|}\right]^{\gamma^2}\left[
{\prod_{k=1}^{N-1}|\alpha_k|\over|\alpha_N|}\right]^{-\gamma^2/2}
\left[{|\alpha_n|\over|\alpha_N|}\right]^{4\gamma^2}
\eea
Note that this formula embraces the previously obtained
special case $n=N$.

Finally we combine this correction factor with the
measure, setting $Z_{N-1}=1$:
\bea
C\left|{\partial T\over\partial Z}\right| 
{\det}^{-(D-2)/2}(-\nabla^2)
&=&\epsilon^{(N-2)\gamma^2/2}\prod_{k=1}^N{1\over \sqrt{|
\alpha_k|}}\nonumber\\
&&\hskip-4.5cm 
\left[{|\alpha_n|\over|\alpha_N|}\right]^{4\gamma^2}
\left[{\prod_{k<N}|\alpha_k|\over
|\alpha_N|}\right]^{(26-D)/16-\gamma^2/2}\left[{\prod_{r<t}|x_t-x_r|
\prod_{m<l<N}|Z_l-Z_m|
\over \prod_{r,\ l<N}|Z_l-x_r|}\right]^{(26-D)/24+\gamma^2}\nonumber\\
&=&\epsilon^{(N-2)\gamma^2/2}\prod_{k=1}^N{1\over \sqrt{|
\alpha_k|}}\nonumber\\
&&\hskip-4.5cm 
\left[{\prod_{k\neq n}|\alpha_k|\over
|\alpha_n|}\right]^{(26-D)/16-\gamma^2/2}\left[{|\alpha_n|^3
\over|\alpha_N|^3}{\prod_{r<t}|x_t-x_r|
\prod_{m<l<N}|Z_l-Z_m|
\over \prod_{r,\ l<N}|Z_l-x_r|}\right]^{(26-D)/24+\gamma^2}
\eea
We recall that the $N$ point tree amplitude is obtained by
multiplying this measure factor by the factor
\bea
dZ_2\cdots dZ_{N-2}\prod_{m<l<N}|Z_l-Z_m|^{2p_l\cdot p_m},\qquad p_k^2
={\bfs p}^2-2p^+p^-={D-2\over24}-\gamma^2
\eea
and integrating the $Z$'s over the range 
$Z_1=0<Z_2<Z_3<\cdots<Z_{N-2}<Z_{N-1}=1$, where $Z_1=0,Z_{N-1}=1,Z_N=\infty$
are held fixed.

It is of interest to write the formula for the scattering amplitude
in a general projective frame where $Z_1<Z_{N-1}<Z_N$ are fixed
to general values. This is done by making a change of variables
by a projective transformation $Z_k\to Y_k=(aZ_k+b)/(cZ_k+d)$ under
which $Z_1=0\to Y_1=b/d$, $Z_{N-1}=1\to Y_{N-1}=(a+b)/c+d)$, 
and $Z_N=\infty\to Y_N= a/c$. In this case the
map from the $z$-plane to the lightcone diagram includes all $N$
terms:
\bea
\rho&=&\sum_{k=1}^N\alpha_k\ln(z-Y_k),\qquad {d\rho\over dz}
={\sum_k\alpha_k\prod_{l\neq k}(z-Y_l)\over\prod_k(z-Y_k)}
\eea
The numerator of $d\rho/dz$ is a polynomial of degree $N-2$
because $\sum_k\alpha_k=0$. Let its roots be $\xi_r$ which are
the images of $x_r$ under the projective transformation
$\xi_r=(ax_r+b)/(cx_r+d)$.
\bea
\sum_k\alpha_k\prod_{l\neq k}(z-Y_l)&=&\left(\sum_l\alpha_l Y_l\right)
\prod_{r=1}^{N-2}(z-\xi_r)\nonumber\\
\alpha_m\prod_{l\neq m}(Y_m-Y_l)&=&\left(\sum_l\alpha_l Y_l\right)
\prod_{r=1}^{N-2}(Y_m-\xi_r)
\eea
The scattering amplitude then becomes
\bea
A_N&=&(Y_N-Y_{N-1})(Y_N-Y_1)(Y_{N-1}-Y_1)\int dY_2\cdots dY_{N-2}
\prod_{m<l}|Y_l-Y_m|^{2p_l\cdot p_m}{\cal M}\\
{\cal M}&=&\epsilon^{(N-2)\gamma^2/2}\prod_{k=1}^N{1\over \sqrt{|
\alpha_k|}} 
\left[{\prod_{k\neq n}|\alpha_k|\over
|\alpha_n|}\right]^{(26-D)/16-\gamma^2/2}\nonumber\\
&&\hskip1in\left[{|\alpha_n|^3
\over|\sum_l \alpha_lY_l|^3}{\prod_{r<t}|\xi_t-\xi_r|
\prod_{m<l}|Y_l-Y_m|
\over \prod_{r,\ l}|Y_l-\xi_r|}\right]^{(26-D)/24+\gamma^2}
\eea
where now factors involving $Y_N$ are included in the various products.

We see that since there are two noncovariant factors raised
to different powers, the only Lorentz covariant choice is
$\gamma=0, D=26$. For $D<26$ the best one can do is either remove the
noncovariant $\xi_r$ dependence by setting $\gamma^2=(D-26)/24$
or remove the other factor with only $\alpha$ dependence by setting
$\gamma^2=(26-D)/8$. We have already seen that in the
first case the external states have ${\bfs p}^2-2p^+p^-
=1$. Then the amplitude is just the generalized $N$-point
Veneziano amplitude times the noncovariant function of the $\alpha$
\bea
\left[{\prod_{k\neq n}|\alpha_k|\over
|\alpha_n|}\right]^{(26-D)/16-\gamma^2/2}&\to&
\left[{\prod_{k\neq n}|\alpha_k|\over|\alpha_n|}\right]^{(26-D)/12}
\eea
This noncovariant factor distinguishes the particle $n$ which 
is assigned $+\gamma$ from the $N-1$ others assigned $-\gamma$,
and it is the only feature that does so. It is interesting that
this factor satisfies tree factorization by itself. This means that
removing it by hand leaves a covariant amplitude that distinguishes
none of the particles and that factorizes as unitarity demands.
This ad hoc procedure would however destroy a local lightcone
worldsheet description,

The second choice $\gamma^2=(26-D)/8$  maintains the scaling behavior demanded
by Lorentz invariance, but sacrifices Lorentz invariance in the
behavior of excited states. In this case the external state momenta
satisfy
\bea
{\bfs p}^2-2p^+p^-&=&{D-2\over24}-{26-D\over8}={D-20\over6}
\eea
The $N$-point scattering amplitude is then proportional to
\bea
A_N&=&\int dZ_2\cdots dZ_{N-2}\prod_{k<l<N}|Z_l-Z_k|^{2p_k\cdot p_l}
 \left[{|\alpha_n|^3\over|\alpha_N|^3}{\prod_{r<t}|x_t-x_r|
\prod_{m<l<N}|Z_l-Z_m|
\over \prod_{l,r}|Z_l-x_r|}\right]^{(26-D)/6}\nonumber\\
&=&\int dZ_2\cdots dZ_{N-2}\prod_{k<l<N}|Z_l-Z_k|^{2p_k\cdot p_l}
 \left[{|\alpha_N|^{N-3} |\alpha_n|^3\over
\prod_k|\alpha_k|}{\prod_{r<t}|x_t-x_r|
\over \prod_{k<l<N}|Z_l-Z_k|}\right]^{(26-D)/6}\nonumber\\
&=&\int dZ_2\cdots dZ_{N-2}\prod_{k<l<N}|Z_l-Z_k|^{2p_k\cdot p_l-(26-D)/6}
 \left[{|\alpha_N|^{N-3} |\alpha_n|^3\over
\prod_k|\alpha_k|}\prod_{r<t}|x_t-x_r|\right]^{(26-D)/6}\nonumber
\eea
Note that the role played by the field $\phi$ in this discussion is
similar to that of the Liouville field in Polyakov's treatment of
the subcritical string \cite{chodost,polyakov,curtrightthorn,gervaisneveu}.
\subsection{Four Point Examples}
We have seen that some aspect of Lorentz invariance is lost
when $D<26$. To illustrate this we work out the 4-point amplitude
in various cases. We first look at the unmodified lightcone
4-point amplitude at general $D$ (taking $Z_1=0,Z_2=Z,Z_3=1,Z_4=\infty$):
\bea
A_4&=&\int dZ\prod_{k=1}^4{1\over |
\alpha_k|^{(D-2)/48}}\left[|\alpha_4|{|x_2-x_1|
\over Z(1-Z)}\right]^{(26-D)/24}
Z^{2p_1\cdot p_2}(1-Z)^{2p_2\cdot p_3}\nonumber\\
&=&\int dZ\prod_{k=1}^4{1\over |
\alpha_k|^{(D-2)/48}}\left[|\alpha_4||x_2-x_1|\right]^{(26-D)/24}
Z^{-\alpha(s)-1}(1-Z)^{-\alpha(t)-1}
\eea 
Let us define 
\bea
2p_1\cdot p_2-(26-D)/24
&=&(p_1+p_2)^2-2{(D-2)/24}-(26-D)/24\nonumber\\
&\equiv& -s-(D-2)/24 -1\equiv -\alpha(s)-1\\
 2p_2\cdot p_3-(26-D)/24
&\equiv& -\alpha(t)-1\\
 \alpha_{ij}&\equiv&\alpha_i+\alpha_j.
\eea
Then
\bea
|\alpha_4|^2|x_2-x_1|^2&=&(\alpha_1+\alpha_2+Z(\alpha_1+\alpha_3))^2
+4Z\alpha_1\alpha_4\nonumber\\
&=&\alpha_{12}^2(1-Z)^2+\alpha_{23}^2Z^2
+(\alpha_{12}^2+\alpha_{23}^2-\alpha_{13}^2)Z(1-Z)
\eea
The last form shows that, in spite of the lack of Lorentz invariance,
$A_4$ is crossing symmetric, which immediately follows from the change
of variables $Z\to1-Z$. To check factorization, note that the
poles in $s$ arise from $Z\sim0$, for which
\bea
|\alpha_4|^2|x_2-x_1|^2&\to&(\alpha_1+\alpha_2)^2
\eea
so the contribution of this factor to the residue is
$|\alpha_{12}|^{1-(D-2)/24}$. Thus we have
\bea
A_4&\sim&\prod_{k=1}^4{1\over |\alpha_k|^{(D-2)/48}}{|\alpha_{12}|^{1-(D-2)/24}
\over{\bfs p}^2-(D-2)/24-|\alpha_{12}|p^-}\nonumber\\
&\sim&|\alpha_1\alpha_2\alpha_{12}|^{-(D-2)/48}
{1\over \left[{\bfs p}^2-(D-2)/24\right]/|\alpha_{12}|-p^-}
|\alpha_3\alpha_4\alpha_{34}|^{-(D-2)/48}\eea
which is precisely the desired factorization property. In this
way we see that the unmodified scattering amplitudes for $D<26$
are crossing symmetric (cyclic), and unitary (factorizing
poles), but lack Lorentz invariance because of the $p^{+}_k$
dependence.
\subsection{Branion Branion Scattering}
\begin{figure}[ht]
\begin{center}
\includegraphics[width=4in]{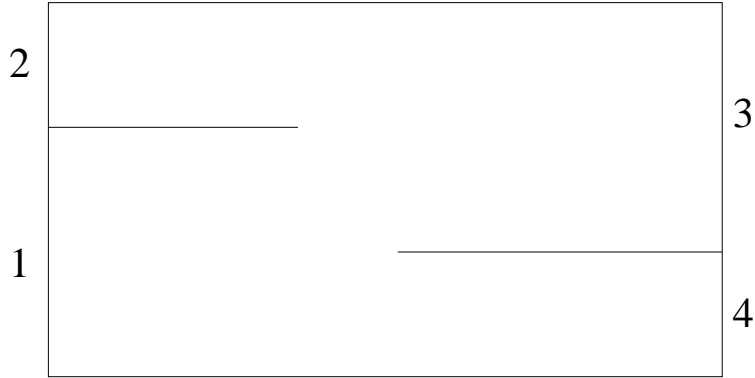}
\caption{Worldsheet for branion scattering}
\label{branions}
\end{center}
\end{figure}

In a four dimensional theory the transverse space is two dimensional.
To describe this situation with critical 26 dimensional open strings we
make a 2+22 split of the 24 transverse coordinates $(x^1,x^2;y^1,
\cdots,y^{22})$
and impose Dirichlet conditions $y^a=0$ at both ends of each open string
\cite{dailp}.
The concept of branions was introduced in the context
of quantum field theory \cite{rozowskyt} to get a handle on the
force between external sources in lightcone quantization, where
it is beneficial to maintain $p^+$ conservation: the sources
are fixed in transverse space but free to move in the longitudinal
direction.  
The physical situation of branion-branion scattering in string
theory 
is an open string, with one end free to move in the two-dimensional
${\bfs x}$ space and the other end fixed at say
${\bfs x}=0$, scattering from 
another open string, also with one free end and
the other end fixed to a different point, say ${\bfs x}={\bfs R}$. 
Only the free ends participate in the interactions (see Fig.~\ref{branions}).
Examples of such string amplitudes have been obtained long ago in \cite{siegel}
in the context of building dual resonance amplitudes with Regge
trajectories with intercepts less than 1.

It is just as easy to analyze a $D=d+2$ dimensional theory using a $(d,24-d)$
slit of transverse space. The ground state mass of each branion
string is then given by
\bea
\alpha^\prime m_{\rm branion}^2=-{24-d\over24}+{d\over48}={d-16\over16}.
\eea
Because the scattering kinematics is 1+1 dimensional, $p^\pm$ conservation
implies either forward or backward scattering. The figure shows
backward scattering $p_4^+=-p_2^+$ and $p_3^+=-p_1^+$, with corresponding 
relations among the $\alpha_k$'s. In the mapping to the upper half
plane we choose $Z_1=1$, $Z_2=U$, $Z_3=0$, $Z_4=\infty$.  Then the
scattering amplitude is
\bea
{\cal M}&=&{g^2\over4p_1^+p_2^+}\int_0^1 dU{\cal D}(U)^{d}
q^{T_0R^2/4\pi}
(1-U)^{-(d-16)(p_1^+/p_2^++p_2^+/p_1^+)/16}
U^{(d-16)(p_1^+/p_2^++p_2^+/p_1^+)/16}\nonumber\\
&=&{g^2\over4p_1^+p_2^+}\int_0^1 dU{\cal D}(U)^{d}q^{T_0R^2/4\pi}
(1-U)^{2p_1\cdot p_2}
U^{-2p_1\cdot p_2}\nonumber\\
&=&{g^2\over4p_1^+p_2^+}\int_0^1 
{4(1-k)dk\over(1+k)^3}q^{T_0R^2/4\pi}
\left({4k\over(1-k)^2}\right)^{2p_1\cdot p_2}{\cal D}(k)^{d}
\label{branionamp}
\eea
here $q=e^{-\pi K^\prime/K}$ is the modulus associated with the map of
the rectangle to the upper half plane (see Fig.~\ref{rectangle}).
The $R$ dependence arises after shifting the string coordinates
by the classical solution, ${\bfs x}_c={\bfs R}(K-x)/2K$ 
in the $z=x+iy$ plane,
that sets both Dirichlet boundary conditions to ${\bfs x}=0$.  
In the discussion of that figure, we established that $U=(1-k)^2/(1+k)^2$,
which was used to obtain the last line. 

Consulting (\ref{cald}) in Appendix B, we have
\bea
{\cal D}^{24}&=& {\det}_{\rm DNDN}^{-12}{(1+k)^2\over4k^2(1-k)^4}
=(2K)^{-6}q^{-1/2}\prod_m(1-q^m)^{-12}{(1+k)^2\over4k^2(1-k)^4}\\
{\cal M}&=&{g^2\over p_1^+p_2^+}\int_0^1 
{(1+k)^{(d-16)/4}dk\over (2k)^{d/12}(1-k^2)^{(d-6)/6}}
\left({4k\over(1-k)^2}\right)^{2\alpha^\prime
p_1\cdot p_2}
{q^{T_0R^2/4\pi-d/48}\over(2K)^{d/4}\prod_m(1-q^m)^{d/2}}
\eea
We recall the relations between $k,K$ and $q$:
\bea
k^2&=&{\theta_2(0)^4\over\theta_3(0)^4}=16q\prod_{n=1}^\infty{(1+q^{2n})^8\over
(1+q^{2n-1})^8}\\
1-k^2&=&{\theta_4(0)^4\over\theta_3(0)^4}=
\prod_{n=1}^\infty{(1-q^{2n-1})^8\over
(1+q^{2n-1})^8}\\
(2K)^2&=&\pi^2{\theta_3(0)^4}={\pi^2}\prod_{n=1}^\infty(1+q^{2n})^8
(1-q^{2n})^4
\eea
From these relations we see that $q\sim k^2/16$ for $k\to0$. In this limit
the integrand then behaves as
\bea
{dk\over(2k)^{d/12}}\left({4k}\right)^{2\alpha^\prime
p_1\cdot p_2}
{k^{T_0R^2/2\pi-d/24}\over\pi^{d/4}}={{4}^{2\alpha^\prime
p_1\cdot p_2}\over(2\pi^3)^{d/12}}{dk}
{k^{T_0R^2/2\pi+2\alpha^\prime
p_1\cdot p_2-d/8}}
\eea
We note that the invariant (mass)$^2$ in the 12 channel $-M^2=(p_1+p_2)^2
=2p_1\cdot p_2-(d-16)/8\alpha^\prime$ since $\alpha^\prime p_k^2=-(d-16)/16$. Integration near $k=0$
then generates a pole at
\bea
\alpha^\prime M^2={T_0R^2\over2\pi}-1\qquad{\rm or}\qquad
M^2=T_0^2 R^2-{1\over\alpha^\prime}
\eea
since $\alpha^\prime=1/2\pi T_0$. This is in accord with the presence
of a stretched string of mass $M\sim T_0R$ 
between ${\bfs x}=0$ and ${\bfs x}={\bfs R}$ in the 12 channel.
The zero point energy squared $-1/\alpha^\prime$ is also in accord
with that of a DD ground string.

Singularities in the 23 channel arise from integrating near $k=1$.
To analyze them it is best to do a Jacobi transformation on
the various infinite products. Define $\bar{q}$ via the relation
$\ln q \ln\bar{q}=\pi^2$, so $q\to1$ implies $\bar{q}\to0$. Then
\bea
\left({-\ln\bar{q}\over\pi}\right)^{1/2}\bar{q}^{1/12}\prod_{k=1}^\infty(1-\bar{q}^{2k})&=&
{q}^{1/12}\prod_{k=1}^\infty(1-{q}^{2k})\\ 
\bar{q}^{-1/24}\prod_{k=1}^\infty(1+\bar{q}^{2k-1})&=&
{q}^{-1/24}\prod_{k=1}^\infty(1+{q}^{2k-1})\\ 
\bar{q}^{-1/24}\prod_{k=1}^\infty(1-\bar{q}^{2k-1})&=&
2^{1/2}{q}^{1/12}\prod_{k=1}^\infty(1+{q}^{2k})\\
2^{1/2}\bar{q}^{1/12}\prod_{k=1}^\infty(1+\bar{q}^{2k})&=&
{q}^{-1/24}\prod_{k=1}^\infty(1-{q}^{2k-1})\eea
From these identities we infer
\bea
k^2&=&\prod_{n=1}^\infty{(1-\bar{q}^{2n-1})^8\over
(1+\bar{q}^{2n-1})^8},\qquad 1-k^2=
16\bar{q}\prod_{n=1}^\infty{(1+\bar{q}^{2n})^8\over
(1+\bar{q}^{2n-1})^8}\\
(2K)^2&=&\pi^2{\theta_3(0)^4}=
{\pi^2}\left({-\ln\bar{q}\over\pi}\right)^{2}\prod_{n=1}^\infty(1+\bar{q}^{2n})^8(1-\bar{q}^{2n})^4\\
\prod_m(1-q^m)&=&\prod_k(1-q^{2k})(1-q^{2k-1})\nonumber\\
&=&\bar{q}^{1/6}q^{-1/24}\left({-\ln\bar{q}\over\pi}\right)^{1/2}\sqrt{2}
\prod_k(1-\bar{q}^{4k})\\
(2K)^{d/4}\prod_m(1-q^m)^{d/2}&=&
{(2\pi)^{d/4}}\left({-\ln\bar{q}\over\pi}\right)^{d/2}
\prod_{n=1}^\infty(1+\bar{q}^{2n})^{d/2}(1-\bar{q}^{4n})^{d}
\bar{q}^{d/12}q^{-d/48}
\eea
Thus we see that $\bar{q}\to0$ implies that $k\to1$. To analyze
integration near $k=1$, we substitute these relations in the
integrand, dropping terms that vanish like a power of $(1-k)$ or
a power of $\bar{q}$:
\bea
{\cal I}&\sim&{dk\over 8(1-k)^{(d-6)/6}}\left({4\over(1-k)^{2}}
\right)^{2\alpha^\prime p_1\cdot p_2}
{q^{T_0R^2/4\pi}\over{(2\pi)^{d/4}}
\bar{q}^{d/12}}\left({-\ln\bar{q}\over\pi}\right)^{-d/2}\nonumber\\
&\sim&{d\bar{q}\over{2(2\pi)^{d/4}}\bar{q}}
\left({1\over16\bar{q}^2}\right)^{2\alpha^\prime
p_1\cdot p_2+(d-8)/8}
\left({-\pi\over\ln\bar{q}}\right)^{d/2}{e^{\pi T_0R^2/4\ln\bar{q}}}
\eea
Integration near $\bar{q}=0$ generates a branch point (because
of the powers of $\ln\bar{q}$) in the variable
$(p_2+p_3)^2=(p_1-p_2)^2=-2p_1\cdot p_2-(d-16)/8\alpha^\prime$
at $1/\alpha^\prime$. This precisely reflects the propagation of the
open string tachyon (with (mass)$^2=-1/\alpha^\prime)$ 
between two points in transverse space.

The small $\bar{q}$ region of integration controls the large $R$
behavior of the scattering amplitude. To clarify this point, it
is helpful to change variables to $T=-\ln\bar{q}$ which is large for small
$\bar{q}$. Then we apply a saddle point approximation to the integral
\bea
I&=&\int_\Lambda^\infty dT\left({\pi\over T}\right)^{d/2}
\exp\left\{-2(\alpha^\prime p_{23}^2-1)T-{\pi T_0R^2\over4T}\right\}\nonumber\\
&\approx&{\pi^{3/4}(T_0R^2)^{1/4}
\over 2^{5/4}(\alpha^\prime p_{23}^2-1)^{3/4}} 
\left({8\pi(\alpha^\prime p_{23}^2-1)\over T_0R^2}\right)^{d/4}
\exp\left\{-R\sqrt{p_{23}^2-1/\alpha^\prime}\right\}\eea
where the saddle point is at $T=\sqrt{\pi T_0 R^2/8(\alpha^\prime p_{23}^2-1)}$
which is large for large $R$. Notice that for 4 dimensional spacetime
($d=2$), this large $R$ behavior is precisely that of the
Kelvin Bessel function $K_0(R\sqrt{p_{23}^2-1/\alpha^\prime})$,
which is the $R$ dependence of the corresponding scattering
amplitude in quantum field theory.

To compare our results to those of Siegel \cite{siegel}, we make use of
some identities from the theory of elliptic functions \cite{bateman}
to rewrite our expression for ${\cal D}$. First note that
\bea
2^{1/6}q^{1/24}\prod_m(1-q^m)&=&\theta_4(0)^{2/3}\theta_2(0)^{1/6}\theta_3(0)^{1/6}
=\sqrt{2K\over\pi}(1-k^2)^{1/6}k^{1/12}
\eea
Then
\bea
{\cal D}^{24}&=&{\pi^6\over16}(2K)^{-12}{1\over4k^3(1-k)^6}
={\pi^6\over16}[2(1+k)K(k)]^{-12}{(1+k)^{12}\over4k^3(1-k)^6}\nonumber\\
&=&\pi^6\left[2K(\sqrt{1-U})\right]^{-12}U^{-3}(1-U)^{-3}
\eea
Then (\ref{branionamp}) can be written
\bea
{\cal M}&=&{g^2\pi^{d/4}\over4p_1^+p_2^+}\int_0^1 dU
\left[2K(\sqrt{1-U})\right]^{-d/2}q^{T_0R^2/4\pi}
(1-U)^{2p_1\cdot p_2-d/8}
U^{-2p_1\cdot p_2-d/8}
\eea
The amplitudes calculated in \cite{siegel} would have $R=0$ and would
not necessarily be for backward scattering. We obtain agreement 
 if we set
$R=0$ in the above formula and $2p_2\cdot p_3=-2p_1\cdot p_2$
in \cite{siegel}.
\vskip14pt
\noindent\underline{Acknowledgments}: 
I should like to thank Warren Siegel and Stuart Dowker for drawing my
attention to their work \cite{siegel,dowker}.
This research was supported in part by the Department of 
Energy under Grant No. DE-FG02-97ER-41029. 

\begin{appendix}
\section{Method of Images}
The empty space solution of the diffusion equation is simply
\bea
P({\bfs\rho}-{\bfs\rho}^\prime,t)={1\over2\pi t}
e^{-({\bfs\rho}-{\bfs\rho}^\prime)^2/2t}
\eea
For a wedge of angle $\theta=\pi/2M$ D or N boundary conditions can be
arranged by placing sources at angles $\pm\alpha+n\pi/M$, $n=0,1\ldots,
2M-1$ where the
source in the wedge is at angle $\alpha$. To impose N conditions on
both edges of the angle choose the same sign for all sources. To
impose Dirichlet conditions on both edges, the image charges alternate
in sign. Finally to arrange D conditions on the abscissa and N
conditions on the ray $\theta=\pi/2M$, choose the sign pattern
$++--++--++\cdots++--$ counterclockwise around the circle. 
Here the first + is the sign of the original source.

For all cases, as one goes counterclockwise around the circle, 
$({\bfs\rho}-{\bfs\rho}_{n\mp})^2$ assumes the values
$2\rho^2(1-\cos(n\pi/M -2\alpha)$, $2\rho^2(1-\cos(n\pi/M)$.
 Then
\bea
\Tr e^{t\nabla^2/2}&=&\int_{\rm wedge} d^2\rho
{1\over2\pi t}\Bigg[1+\sum_{n=1}^{2M}\pmatrix{-\cr+\cr(-)^{n-1}\cr}
e^{-\rho^2[1-\cos(n\pi/M-2\alpha)]/t}\nonumber\\
&&\hskip2in+\sum_{n=1}^{2M-1}
\pmatrix{+\cr+\cr(-)^n\cr}e^{-\rho^2[1-\cos(n\pi/M)]/t} \Bigg]
\eea
where the signs in the sums are for boundary conditions $DD,NN,DN$
respectively.

Except for the 1 term and the $n=1,2M$ terms of the first sum, the
integral of $\rho$ over the whole infinite wedge is finite. The integral
of the 1 term is simply $A/2\pi t$ where $A$ is the area of the wedge.
The integral of the $n=2M$ term of the first sum, 
restricting $0<\rho<R(\alpha)$,
\bea
&&\hskip-28pt{1\over2\pi t}\int_0^{\pi/2M}d\alpha\int_0^R\rho d\rho 
e^{-2(\rho^2/t)\sin^2\alpha}
={1\over4\pi}\int_0^{\pi/2M}d\alpha {1\over2\sin^2\alpha} 
(1-e^{-2(R^2/t)\sin^2\alpha})\nonumber\\
&=&{1\over8\pi}\int_0^{\pi/2M}d\alpha (1-e^{-2(R^2/t)\sin^2\alpha})
{d\over d\alpha}(-\cot\alpha)\nonumber\\
&=&-{1\over8\pi}\cot{\pi\over2M}(1-e^{-2(R^2/t)\sin^2\pi/2M})
+{1\over4\pi t}\int_0^{\pi/2M}d\alpha e^{-2(R^2/t)\sin^2\alpha}\cot\alpha
{d\over d\alpha}(R^2\sin^2\alpha)\nonumber\\
&\sim&-{1\over8\pi}\cot{\pi\over2M}
+{1\over4\pi t}\int_0^{\infty}{d\alpha\over\alpha} e^{-2(\alpha^2R^2/t)}
2(RR^\prime\alpha^2+R^2\alpha)\nonumber\\
&\sim&-{1\over8\pi}\cot{\pi\over2M}
+{1\over2\pi}\int_0^{\infty}{d\eta} e^{-\eta^2}
\left({R^\prime \over2R}\eta+{R\over\sqrt{2t}}\right)
= -{1\over8\pi}\cot{\pi\over2M}
+{R(0)\over4\sqrt{2\pi}}+O\left({R^\prime \over R}\right)\nonumber\\
&\sim& -{1\over8\pi}\cot{\pi\over2M}
+{R(0)\over4\sqrt{2\pi}}
\eea
The  integral in the $n=1$ term of the first sum gives the same result with
$R(\pi/2M)$ in place of $R(0)$.
The remainder of the first sum gives
\bea
&&\hskip-1in{1\over2\pi t}\sum_{n=2}^{2M-1}\pmatrix{-\cr+\cr(-)^{n-1}\cr}
\int_0^{\pi/2M}d\alpha\int_0^\infty\rho d\rho
e^{-\rho^2[1-\cos(n\pi/M-2\alpha)]/t}\nonumber\\
&=&{1\over8\pi}\sum_{n=2}^{2M-1}\pmatrix{-\cr+\cr(-)^{n-1}\cr}
\left[\cot{(n-1)\pi\over2M}-\cot{n\pi\over2M}\right]
\eea
In the first two cases ($DD,NN$) the inner terms in the sum cancel in pairs
leaving the first term for $n=2$ and the second term for $n=2M-1$:
\bea
 {1\over8\pi}\sum_{n=2}^{2M-1}
\left[\cot{(n-1)\pi\over2M}-\cot{n\pi\over2M}\right]&=&
 {1\over8\pi}\left[\cot{\pi\over2M}-\cot{(2M-1)\pi\over2M}\right]\nonumber\\
&=&{1\over4\pi}\cot{\pi\over2M}
\eea
In the last case a complete cancellation occurs ``outside-in'':
\bea
\sum_{n=2}^{2M-1}
(-)^{n-1}\left[\cot{(n-1)\pi\over2M}-\cot{n\pi\over2M}\right]
&=&
\sum_{n=1}^{2M-2}
(-)^{n}\cot{n\pi\over2M}+\sum_{n=2}^{2M-1}
(-)^{n}\cot{n\pi\over2M}\nonumber\\
&=&
\sum_{n=1}^{2M-2}
(-)^{n}\left[\cot{n\pi\over2M}+\cot{(2M-n)\pi\over2M}\right]\nonumber\\
&=&0
\eea
The 1 term together with all the contributions to the first
sum in square brackets contribute simply
\bea
{A\over2\pi t}\pmatrix{-\cr+\cr-\cr}{R(0)\over4\sqrt{2\pi t}}
\pmatrix{-\cr+\cr+\cr}{R(\pi/2M)\over4\sqrt{2\pi t}}&=&
{A\over2\pi t}-{L_D-L_N\over4\sqrt{2\pi t}}
\eea
where $L_D,L_N$ are the total lengths of the Dirichlet and Neumann
boundaries respectively.

Finally we turn to the second sum in square brackets, which will be responsible
for the corner contributions. The $\rho$ integration is finite and
elementary:
\bea
\Tr \{e^{t\nabla^2/2}\}_{\rm corner}=
{1\over16M}\sum_{n=1}^{2M-1}
\pmatrix{+\cr+\cr(-)^n\cr}{1\over\sin^2(n\pi/2M)}=
-{1\over4M}\sum_{n=1}^{2M-1}
\pmatrix{+\cr+\cr(-)^n\cr}{e^{-in\pi/M}\over(1-e^{-in\pi/M})^2}
\eea
This sum can be represented as a contour integral because the
quantities $z_n\equiv e^{-in\pi/M}$ are all the non unit $2M$th
roots of unity: $z_n^{2M}-1=0$ and $z_n^M=(-)^n$. We have
\bea
-{(1,z^M)\over2(z^{2M}-1)(z-1)^2}\sim {(1,(-)^n)
\over16M\sin^2(n\pi/2M)}{1\over z-z_n},
\qquad {\rm as}\quad z\to z_n
\eea
Thus
\bea
\Tr \{e^{t\nabla^2/2}\}_{\rm corner}&=&-\oint_C {dz\over2\pi i}
\pmatrix{1\cr1\cr z^M\cr}{1\over2(z^{2M}-1)(z-1)^2}
\eea
where $C$ is a counterclockwise contour encircling all the $z_n$,
for $n=1,2,\ldots,2M-1$. This contour can be deformed to a clockwise
contour encircling the (triple) pole at $z=1$. Then the integral
is just $(-)$ times the residue of that triple pole. In terms of the
functions 
\bea
f(z)\equiv {z-1\over 2(z^{2M}-1)},
\qquad f_1(z)\equiv{z^M(z-1)\over 2(z^{2M}-1)}
={(z-1)\over 2(z^{M}-z^{-M})}
\eea
these residues are just $f^{\prime\prime}(1)/2$, $f^{\prime\prime}(1)/2$,
and $f_1^{\prime\prime}(1)/2$ respectively.
An efficient way to evaluate these derivatives it to put $g(t)=f(e^t)$,
so that $\dot(g)=e^tf^\prime$ and $\ddot(g)=e^tf^\prime+e^{2t}f^{\prime\prime}$.Then $f^{\prime\prime}(1)=\ddot{g}(0)-\dot{g}(0)$. So we expand $g$
to order $t^2$:
\bea
g(t)&=&{1\over4M}{1+t/2+t^2/6\over1+Mt+2M^2t^2/3}={1\over4M}+\left({1\over8M}
-{1\over4}\right)t+\left({1\over24M}+{M\over12}-{1\over8}\right)t^2+O(t^3)\nonumber\\
{1\over2}f^{\prime\prime}(1)&=&{1\over24M}+{M\over12}-{1\over16M}={M\over12}-{1\over48M}
={1\over24}\left(2M-{1\over2M}\right)
\eea
which confirms the formula for a DD or NN corner of angle $\theta=\pi/2M$.

To handle the ND case we expand
\bea
g_1(t)&=&{1\over4M}{1+t/2+t^2/6\over1+M^2t^2/6}={1\over4M}+{t\over8M}
-{M^2-1\over24M}t^2+O(t^3)\nonumber\\
{1\over2}f^{\prime\prime}(1)&=&-{M\over24}+{1\over24M}-{1\over16M}
=-{M\over24}-{1\over48M}=-{1\over48}\left(2M+{2\over2M}\right)
\eea
which confirms the formula for a DN corner of angle $\theta=\pi/2M$.
\section{Determinant for the Lightcone Worldsheet Tree}

The quantities $Z_k$, with $k=1\cdots (N-1)$, and $x_r$, 
with $r=1\cdots(N-2)$ are determined from the map from the
upper-half Koba-Nielsen plane ($z$) to the lightcone world sheet
($\rho=\tau+i\sigma$):
\bea
\rho&=&\sum_{k=1}^{N-1}\alpha_k\ln(z-Z_k),\qquad {d\rho\over dz}\bigg|_{z=x_r}=0\\
{d\rho\over dz}&=&\sum_{k=1}^{N-1}{\alpha_k\over z-Z_k}
={\sum_{k=1}^{N-1}\alpha_k\prod_{l\neq k}(z-Z_l)\over
\prod_k(z-Z_k)}=-\alpha_N{\prod_r (z-x_r)\over\prod_k(z-Z_k)}\\
{d^2\rho\over dz^2}\bigg|_{z=x_s}&=&\sum_{k=1}^{N-1}{\alpha_k\over z-Z_k}
=-\alpha_N{\prod_{r\neq s} (x_s-x_r)\over\prod_k(x_s-Z_k)}
\eea
where the last line is true because the factor $(z-x_s)$ in the numerator
must be killed by the derivative to get a nonzero contribution.
The asymptotic strings at $\tau=\pm\infty$ are mapped from
the $Z_k$. In this notation $Z_N=\infty, Z_1=0$. 
A useful identity follows by setting $z=Z_m$ in the identity
\bea
-\alpha_N\prod_r (z-x_r)&=&\sum_{k=1}^{N-1}\alpha_k\prod_{l\neq k}(z-Z_l)\\
-\alpha_N\prod_r (Z_m-x_r)&=&\sum_{k=1}^{N-1}\alpha_k\prod_{l\neq k}(Z_m-Z_l)
=\alpha_m\prod_{l\neq m}(Z_m-Z_l)\\
|\alpha_N|^{N}\prod_{m,r}|Z_m-x_r|&=&\prod_{m=1}^{N}|\alpha_m|
\prod_{l\neq k}|Z_k-Z_l|
\eea
We next consider the transformation of the determinant.
\bea
\Sigma=\ln|\alpha_N|-\sum_{k=1}^{N-1}\ln|z-Z_k|+\sum_{r=1}^{N-2}\ln|z-x_r|
\eea
Clearly $\partial_y\Sigma=0$ on the real axis. Since the points
$z=Z_k,x_s$ are singular, we deform the boundary near those points into small
semicircles, in the upper half plane, of radii $\epsilon_k,\epsilon_r$ 
respectively. The radius
$\epsilon_k$ near $Z_k$ can be interpreted in terms of a large time
$T_k$ for the asymptotic string $k$. From the mapping function we find
\bea
\epsilon_k=e^{T_k/\alpha_k}\prod_{l\neq k}|Z_l-Z_k|^{-\alpha_l/\alpha_k}
\label{epsilont}
\eea
The string $N$ is asymptotic at large $z$. If $R$ is the radius of
a large semi-circle, we have from the mapping function
\bea
T_N\sim -\alpha_N\ln R, \qquad R\sim e^{-T_N/\alpha_N}.
\eea
On the other hand the radius $\epsilon_s$ near $x_s$ is a temporary regulator,
which maps onto a circular deformation of the boundary
near the corresponding interaction
point on the lightcone worldsheet. From the mapping function we see that
the radius of this regulating circle on the worldsheet is given
by
\bea
\delta_s&=&{1\over2}\epsilon_s^2\bigg|{d^2\rho\over dz^2}\bigg|_{z=x_s}
={1\over2}\epsilon_s^2|\alpha_N|{\prod_{r\neq s} 
|x_s-x_r|\over\prod_k|x_s-Z_k|}
\\
\epsilon_s&=&\sqrt{2\delta_s\over|\alpha_N|}
{\prod_k|x_s-Z_k|^{1/2}\over\prod_{r\neq s} 
|x_s-x_r|^{1/2}}\\ 
\prod_s\epsilon_s&=&|\alpha_N|^{-N+3/2}\prod_k|\alpha_k|^{1/2}
\prod_s\sqrt{2\delta_s}
{\prod_{l\neq k}|Z_l-Z_k|^{1/2}\over\prod_{r\neq s} 
|x_s-x_r|^{1/2}}
\eea
To calculate the determinant for the lightcone worldsheet, we start
with the determinant for the region in the upper-half $z$-plane bounded by the
real axis, the large radius $R$ semi-circle, and the small radius
$\epsilon_k,\epsilon_r$ semi-circles. Then we apply the
generalized McKean-Singer formula to transform to the determinant 
for the worldsheet.
\subsection{Unmixed Boundary Conditions}
In this case, the boundary conditions are either
Dirichlet everywhere or Neumann everywhere.
Then in the limit of large $R$
and small $\epsilon$, factorization implies that the $z$-plane
figure determinant has the behavior
\bea
-{1\over2}\Tr\ln(-\nabla^2)_z&\sim& {5\over24}\ln R +{1\over24}\sum_k
\ln\epsilon_k+{1\over24}\sum_r\ln\epsilon_r +{\rm const}
\eea
where the constant term, representing the determinant for the
upper half plane with the same boundary conditions everywhere,
has nothing to depend on! 
We treat mixed boundary conditions in the next subsection,
where the corresponding term can depend on the relative locations
of the points that separate Dirichlet from Neumann boundary
conditions.

Next we develop the transformation of the determinant from this
$z$-plane figure to the lightcone worldsheet:
\bea
\Sigma=\ln|\alpha_N|-\sum_{k=1}^{N-1}\ln|z-Z_k|+\sum_{r=1}^{N-2}\ln|z-x_r|
\label{sigmaws}
\eea
Clearly $\partial_y\Sigma=-\partial_n\Sigma=0$ on the real axis. Thus
the change formula receives contributions from the corners
and semi-circles only. For $z$ near $Z_k$ put $z=Z_k+r e^{i\varphi}$
and approximate
\bea
\Sigma\approx\ln|\alpha_N|-\ln r-\sum_{l\neq k}^{N-1}\ln|Z_l-Z_k|
+\sum_{r=1}^{N-2}\ln|Z_k-x_r|,\qquad \partial_n\Sigma\approx {1\over r}
\eea
Then
\bea
\Delta_{\epsilon_k}&=&\left[{1\over24}-{1\over12}+{1\over8}\right]\Sigma
={1\over12}\ln\left({|\alpha_N|\over\epsilon_k}{\prod_r|Z_k-x_r|
\over\prod_{l\neq k}|Z_k-Z_l|}\right)=
\ln\left({|\alpha_k|\over\epsilon_k}\right)^{1/12}
\label{changeNN}
\eea
The three terms in square brackets are the $\int dl\Sigma\partial_n\sigma$
term the extrinsic curvature term (negative here) 
and the two corners at this semi-circle respectively.

For $z=x_s+re^{i\varphi}$, on the other hand we have
\bea
\Sigma\approx\ln|\alpha_N|+\ln r-\sum_{l}^{N-1}\ln|Z_l-x_s|
+\sum_{r\neq s}\ln|x_s-x_r|,\qquad \partial_n\Sigma\approx -{1\over r}
\eea
Then
\bea
\Delta_{\epsilon_s}=\left[-{1\over24}-{1\over12}+{1\over8}\right]\Sigma=0
\label{changex}
\eea
Finally for the large semi-circle, $\Sigma\approx -\ln (r/|\alpha_N|)$,
$\partial_n\Sigma\approx -1/r$, and
\bea
\Delta_R=\left[-{1\over24}+{1\over12}+{1\over8}\right]\Sigma
=-{1\over6}\ln {R\over|\alpha_N|}
\eea
Combining all the contributions,
we have
\bea
{\det}^{-1/2}(-\nabla^2)_\rho&=&{\det}^{-1/2}(-\nabla^2)_z
\left({|\alpha_N|\over R}\right)^{1/6}\prod_k\left({|\alpha_k|\over\epsilon_k}
\right)^{1/12}\nonumber\\
&=&C{|\alpha_N|}^{1/6}R^{1/24}\prod_k\epsilon_k^{-1/24}\prod_r\epsilon_r^{1/24}
\prod_k{|\alpha_k|}^{1/12}\nonumber\\
&=&C{|\alpha_N|}^{1/6}\exp\left\{-\sum_{k=1}^N{T_k\over24\alpha_k}\right\}
\prod_{k\neq l}|Z_k-Z_l|^{\alpha_k/24\alpha_l}
\prod_k{|\alpha_k|}^{1/12}\nonumber\\
&&\left[{\prod_r(2\delta_r)\over|\alpha_N|^{N-2}}
{\prod_k\prod_r|x_r-Z_k|\over\prod_{r\neq s} 
|x_s-x_r|}\right]^{1/48}\nonumber\\
&=&C{|\alpha_N|}^{1/6}\exp\left\{-\sum_{k=1}^N{T_k\over24\alpha_k}\right\}
\prod_{k\neq l}|Z_k-Z_l|^{\alpha_k/24\alpha_l}
\prod_k{|\alpha_k|}^{1/12}\nonumber\\
&&\left[{\prod_r(2\delta_r)\prod_k|\alpha_k|\over|\alpha_N|^{2N-3}}
{\prod_{k\neq l}|Z_l-Z_k|\over\prod_{r\neq s} 
|x_s-x_r|}\right]^{1/48}\nonumber\\
&=&C\prod_{k=1}^N{|\alpha_k|^{1/8}\over|\alpha_k|^{1/48}}
\prod_{k\neq l}|Z_k-Z_l|^{\alpha_k/24\alpha_l}
\nonumber\\
&&\left[{\prod_r\sqrt{2\delta_r}\over|\alpha_N|^{N-3}}
{\prod_{k< l}|Z_l-Z_k|\over\prod_{r< s} 
|x_s-x_r|}\right]^{1/24}\exp\left\{-\sum_{k=1}^N{T_k\over24\alpha_k}\right\}
\eea
If there are $d=D-2$ transverse dimensions this entire factor should be 
raised to the power $d$.

The worldsheet path integral is this determinant factor times a factor
$e^{iW_c}$ which arises from removing boundary data in the path integral
by shifting the ${\bfs x}$ by the classical solution that satisfies those
boundary data. Among other things $e^{iW_c}$ includes factors 
$R^{-{\bfs p}^2}\prod_k\epsilon_k^{{\bfs p}^2}$ in the limit 
that the $-T_k/\alpha_k$ get large.
If $W_c=\sum_{kl}p_k N(\rho_k,\rho_l)p_l$ is 
expressed in terms of a Neumann function, these factors arise from
the diagonal $l=k$ terms. The rest of these diagonal terms, 
combined with the factors $|\alpha_k|^{1/8}$, provide a factor of
the ground string wave function for each external string. 
The $N$ ground string scattering amplitude is
obtained by amputating these ground state wave functions
together with the factors $e^{\sum_k({\bfs p}_k^2-d/24)T_k/\alpha_k}
=e^{\sum_k T_kP^-_k}$
from the path integral and integrating over the interaction times
$\int d\tau_1\cdots d\tau_{N-2}$
where $\rho_r=\tau_r+i\sigma_r$ are the locations of the $N-2$ interaction
points on the worldsheet. By translational invariance in $x^+$
the integrand after amputation will acquire a factor
$e^{a\sum_kP^-_k}$ if all the $\tau_r$ are translated by $a$. This means
that integrating over one of the $\tau_r$ simply produces
a $P^-$ conserving delta function. The coefficient of this delta function
is just the integral over only $N-3$ of the $\tau_r$. Note that
$\sum_k\alpha_k=0$ by the lightcone worldsheet construction and
$\sum_k P_k=0$ when Neumann conditions are chosen for the                
${\bfs x}$ integrals as explained in Section 3 (see (\ref{conserve})).       
\bea
{\cal M}&=&\int d\tau_2\cdots d\tau_{N-2}\left[{\det}^{-d/2}(-\nabla^2)_\rho e^{iW_c}\right]_{\rm amputated}
\eea
where we have set $\tau_1=0$ and understand that $\sum_kP^-_k=0$.

The final result for $[e^{iW_c}]_{\rm amputated}$ includes the off diagonal
terms in its Neumann function representation, together with the 
parts of $\epsilon_k$ that remain after amputating $e^{\sum_k T_kP^-_k}$:
\bea
\left[e^{iW_c}\right]_{\rm amputated}&=&\prod_{k<l}|Z_l-Z_k|^{2{\bfs p}_k\cdot
{\bfs p}_l}\left(\prod_{k\neq l}|Z_k-Z_l|
\right)^{-\alpha_l{\bfs p}_k^2/\alpha_k}\nonumber\\
\left[{\det}^{-d/2}(-\nabla^2)_\rho\right]_{\rm amputated}
&=&C\prod_{k=1}^N{1\over|\alpha_k|^{d/48}}
\prod_{k\neq l}|Z_k-Z_l|^{d\alpha_k/24\alpha_l}
\nonumber\\&&
\qquad\qquad\left[{\prod_r\sqrt{2\delta_r}\over|\alpha_N|^{N-3}}
{\prod_{k< l}|Z_l-Z_k|\over\prod_{r< s} 
|x_s-x_r|}\right]^{d/24}\nonumber\\
\left[{\det}^{-d/2}(-\nabla^2)_\rho e^{iW_c}\right]_{\rm amputated}
&=&C\prod_{k=1}^N{1\over|\alpha_k|^{d/48}}
\prod_{k<l}|Z_k-Z_l|^{2p_k\cdot p_l}
\nonumber\\&&
\qquad\qquad\left[{\prod_r\sqrt{2\delta_r}\over|\alpha_N|^{N-3}}
{\prod_{k< l}|Z_l-Z_k|\over\prod_{r< s} 
|x_s-x_r|}\right]^{d/24}
\eea
where we have used $p_k\cdot p_l={\bfs p}_k\cdot {\bfs p}_l
-p^+_kp^-_l-p^-_kp^+_l={\bfs p}_k\cdot {\bfs p}_l
-\alpha_k({\bfs p}^2_l-d/24)/2\alpha_l-\alpha_l({\bfs p}^2_k-d/24)/2\alpha_k$
It is convenient to change integration variables from the $\tau$'s to the
$Z$'s. Mandelstam's result for the Jacobian is (taking $Z_1,Z_{N-1},Z_N=0,1,
\infty$ respectively)
\bea
{\partial(\tau_2,\cdots,\tau_{N-2})\over
\partial(Z_2,\cdots,Z_{N-2})}&=&\left[{1
\over|\alpha_N|^{N-3}}
{\prod_{k< l}|Z_l-Z_k|\over\prod_{r< s} 
|x_s-x_r|}\right]^{-1},
\eea
so that the scattering amplitude becomes
\bea
{\cal M}&=&C\prod_r(2\delta_r)^{d/48}\prod_{k=1}^N{1\over|\alpha_k|^{d/48}}
\int dZ_2\cdots dZ_{N-2}
\prod_{k<l}|Z_k-Z_l|^{2p_k\cdot p_l}
\nonumber\\&&
\qquad\qquad\left[{1\over|\alpha_N|^{N-3}}
{\prod_{k< l}|Z_l-Z_k|\over\prod_{r< s} 
|x_s-x_r|}\right]^{(D-26)/24}
\eea
The factor raised to the power $D-26$ depends on the Lorentz frames
so the critical dimension $D=26$ is necessary for Lorentz invariance
\cite{goddardgrt}, in which case ${\cal M}$ is proportional
to the $N$ particle dual resonance amplitude.
Of course factorization implies that $C=g^{N-2}$
and $\delta_r=\delta$, independent of $r$. Then 
$\prod_r(2\delta_r)=(2\delta)^{N-2}$
so $\delta$ can be absorbed in the coupling constant.
\subsection{Mixed Boundary Conditions}
Here we consider the cases where the boundary consists of several segments
with either Dirichlet or Neumann boundary. Call the points that separate
different boundary conditions $P_a$. The asymptotic strings on
the world sheet can now have two free ends (NN), one free end (ND),
or no free ends (DD). It will be convenient to choose to close the
asymptotic world sheet with N, D, and D boundary conditions
respectively, in order to minimize the number of ND corners.

The contributions associated with the NN and DD asymptotic strings will
therefore be exactly as in the previous subsection, since they involve
no ND corners. Also the contributions associated with the
interaction points will be the same. Only the contributions from
the ND strings need modification. Since we have to have an even
number of ND strings, in this section we might as well assume there
are at least 2 and take one of them to map to the $z=\infty$.
Then by factorization the $z$-plane determinant has the behavior
for large $R$ and small $\epsilon$
\bea
-{1\over2}\Tr\ln(-\nabla^2)_z&\sim& {1\over48}\ln R
-{1\over48}\sum_{k\in{\rm DN}}\ln\epsilon_k 
+{1\over24}\sum_{k\in{\rm NN}}
\ln\epsilon_k+{1\over24}\sum_r\ln\epsilon_r + \ln{\cal D}
\eea
where in the last term ${\cal D}(P_a)$, representing
the determinant for the $z$-plane stripped of the semi-circles,
can now depend on the locations of the
Dirichlet-Neumann transitions points $P_a$.

The transform to the worldsheet involves the same $\Sigma$ (\ref{sigmaws}),
the same change factors associated with $x_r$ (\ref{changex}) 
and $Z_k$ for $k\in{\rm NN}$ (\ref{changeNN}) as in the previous subsection.
Modifications occur in the change factor associated with $R$
\bea
\Delta^{\rm DN}_R=\left[-{1\over24}+{1\over12}+0\right]\Sigma
=-{1\over24}\ln {R\over|\alpha_N|}
\eea
and in the change factor associated with $Z_k$ with $k\in{\rm DN}$.
\bea
\Delta^{\rm DN}_{\epsilon_k}&=&\left[{1\over24}-{1\over12}+0\right]\Sigma
=-{1\over24}
\ln\left({|\alpha_k|\over\epsilon_k}\right)
\label{changeDN}
\eea
Combining all the contributions,
we have for the determinant on the worldsheet:
\bea
{\det}^{-1/2}(-\nabla^2)_\rho&=&{\det}^{-1/2}(-\nabla^2)_z
\left({|\alpha_N|\over R}\right)^{1/24}\prod_{k\in{\rm NN}}
\left({|\alpha_k|\over\epsilon_k}\right)^{1/12}\prod_{k\in{\rm DN}}
\left({|\alpha_k|\over\epsilon_k}\right)^{-1/24}\nonumber\\
&=&{\cal D}{|\alpha_N|}^{1/24}R^{-1/48}
\prod_{k\in{\rm NN}}\epsilon_k^{-1/24}
\prod_{k\in{\rm DN}}\epsilon_k^{1/48}
\prod_{k\in{\rm NN}}{|\alpha_k|}^{1/12}\prod_{k\in{\rm DN}}{|\alpha_k|}^{-1/24}
\prod_r\epsilon_r^{1/24}\nonumber\\
&=&{\cal D}{|\alpha_N|}^{1/24}R^{-1/48}
\prod_{k\in{\rm NN}}\epsilon_k^{-1/24}
\prod_{k\in{\rm DN}}\epsilon_k^{1/48}
\prod_{k\in{\rm NN}}{|\alpha_k|}^{1/12}\prod_{k\in{\rm DN}}{|\alpha_k|}^{-1/24}
\nonumber\\
&&\left[{\prod_r(2\delta_r)\prod_k|\alpha_k|\over|\alpha_N|^{2N-3}}
{\prod_{k\neq l}|Z_l-Z_k|\over\prod_{r\neq s} 
|x_s-x_r|}\right]^{1/48}\nonumber\\
&=&{\cal D} R^{-1/48}
\prod_{k\in{\rm NN}}\epsilon_k^{-1/24}
\prod_{k\in{\rm DN}}\epsilon_k^{1/48}
\prod_{k\in{\rm NN}}{|\alpha_k|}^{1/8}\prod_{k=1}^N{|\alpha_k|}^{-1/48}
\nonumber\\
&&\left[{\prod_r\sqrt{2\delta_r}\over|\alpha_N|^{N-3}}
{\prod_{k<l}|Z_l-Z_k|\over\prod_{r< s} 
|x_s-x_r|}\right]^{1/24}\nonumber\\
\eea
Remembering (\ref{epsilont}) we see that the different powers of
$\epsilon_k$ for the NN and DN cases simply reflect the different
ground state masses for the open string in those cases
\bea
\alpha^\prime M_G^2=-{d_{\rm NN}\over24}+{d_{\rm DN}\over48}
\eea
where $d_{\rm NN(DN)}$ is the dimension of NN(DN) string coordinates.
Each NN external string can carry a momentum, so we we collect them
as the components of a $d_{\rm NN}$ dimensional vector ${\bfs p}$.
Then the $p^-$ of the kth string is $p_k^-=({\bfs p}^2+M_G^2)/2p^+_k
=({\bfs p}^2+M_G^2)/\alpha_k$. Then
\bea
\epsilon_k^{{\bfs p}_k^2-d_{\rm NN}/24+d_{\rm DN}/48}&=&e^{T_kp^-_k}
\prod_{l\neq k}|Z_k-Z_l|^{-2\alpha^\prime p^+_lp^-_k}\nonumber\\
&=&e^{T_kp^-_k}
\prod_{l<k}|Z_k-Z_l|^{-2\alpha^\prime p^+_lp^-_k-2\alpha^\prime p^+_kp^-_l}
\eea
The extra factor of $|\alpha_k|^{d_{\rm NN}/8}$ for $k\in{\rm NN}$ 
simply reflects
the normalization of the NN ground state compared to the DN ground state.
\subsection{Determining ${\cal D}$}
We now consider the dependence of ${\cal D}$ 
on the DN transition points. We will content ourselves with working
out that dependence for no more than 2 Dirichlet boundaries (i.e.
no more than 4 DN transition points. For the case of only one
Dirichlet boundary, the two transition points can be taken to
be two of the fixed Koba-Nielsen variables, and ${\cal D}$ will
therefore not depend on any of the integration variables.
For the case of two Dirichlet boundaries there are four transition
points, three of which can be taken fixed, but ${\cal D}$ can
depend on the fourth, which will be an integration variable.

\begin{figure}[ht]
\begin{center}
\includegraphics[width=2.5in]{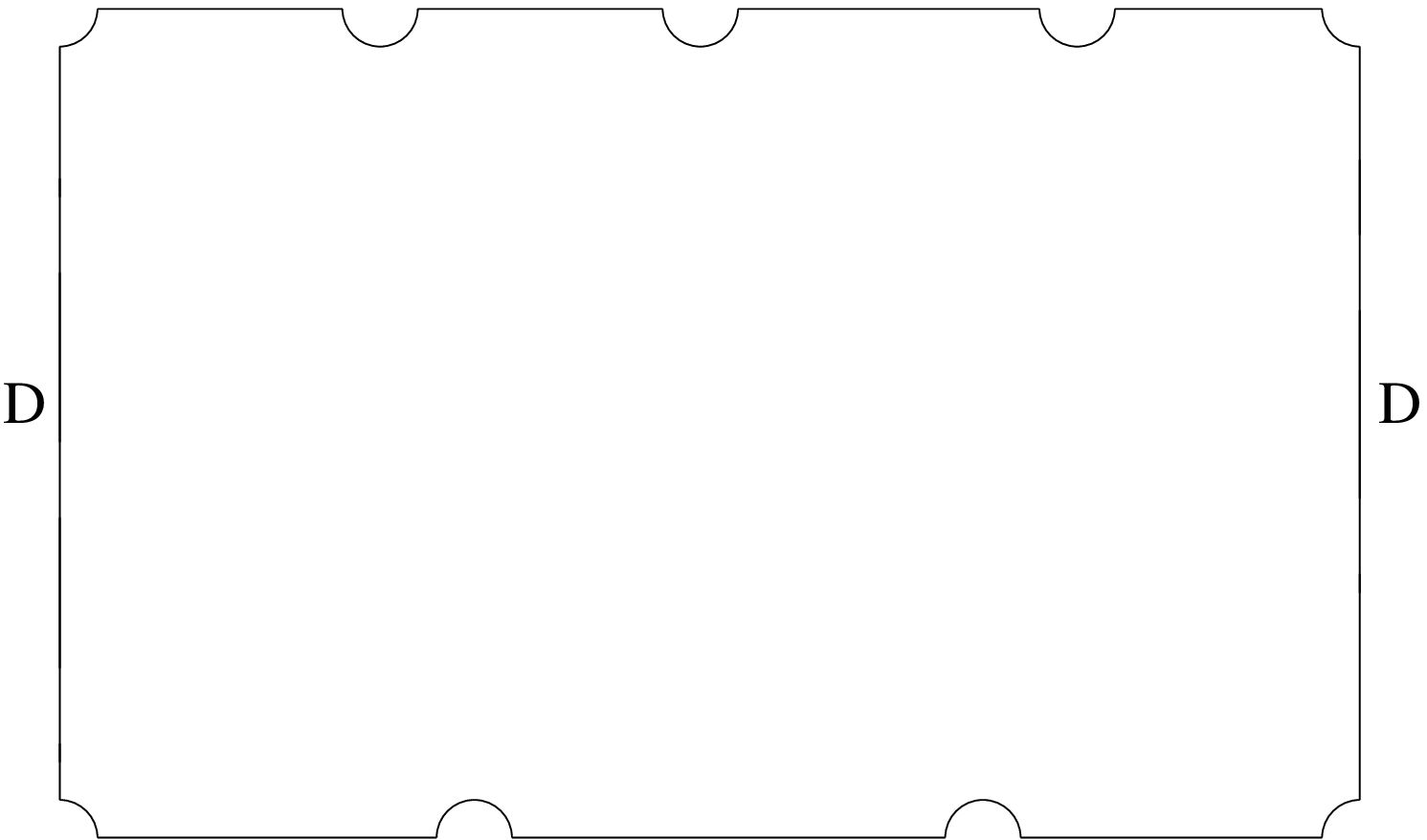}\qquad\qquad
\includegraphics[width=3.0in]{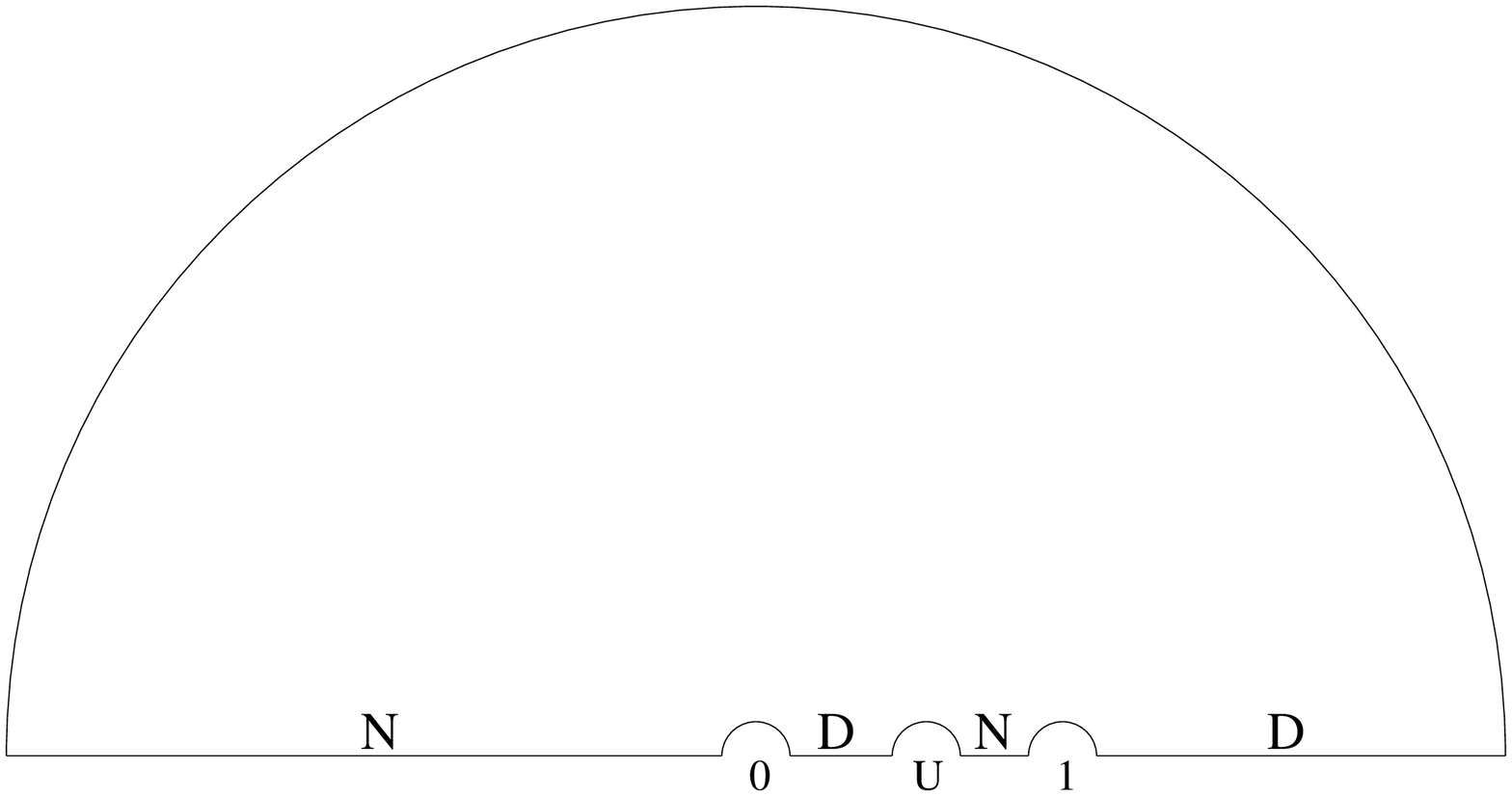}
\caption{Rectangle mapped to the upper half plane by an elliptic function. 
The horizontal boundaries
can be taken Neumann, and the vertical ones can be taken
Dirichlet. The small circular arcs in both figures are all
meant to be infinitesimal and are centered on potentially singular
points of the mapping. The large semicircle on the right is meant to
be infinite.}
\label{rectangle}
\end{center}
\end{figure}

To calculate ${\cal D}$ for the case of two Dirichlet boundaries
we consider the conformal map of a DNDN rectangle to the upper half
plane (see Fig.~(\ref{rectangle}). If the lightcone worldsheet
is mapped to that figure, the asymptotic strings would be mapped
to the centers of the circular arcs on the vertical boundaries.
For the purpose of calculating ${\cal D}$ we only need to keep
the quarter circles at the corners. We map their centers to the
points $0,U,1,\infty$, labelled counterclockwise starting at the
upper left corner. Situate the rectangle in the upper half $z$-plane
with the bottom side on the real axis, $-K<x<+K$, with the upper
boundary on the line $z=x+iK^\prime$. Let $u$ be the complex variable 
of the target upper half plane. Then
\bea
u&=&{(k{\rm\ sn}(z,q)+1)(k-1)\over(k{\rm\ sn}(z,q)-1)(k+1)},\qquad
{du\over dz}={-2k(k-1){\rm sn}^\prime(z,q)\over(k+1)(k{\rm\ sn}(z,q)-1))^2}
\eea 
where ${\rm\ sn}$ is one of the Jacobian elliptic functions of modulus
$k$ and $q=e^{-\pi K^\prime/K}$. With this notation,
$U=(k-1)^2/(k+1)^2$.

The determinant for the $u$-plane figure is in the limit ${\bar\epsilon}_{1,2,3}\to0$ and $R\to\infty$
\bea
-{1\over2}\ln{\det}_u&\sim&{1\over48}\ln R-{1\over48}\ln{\bar\epsilon}_1
{\bar\epsilon}_2
{\bar\epsilon}_3+\ln{\cal D}
\eea
where $R$ is the radius of the large semicircle and ${\bar\epsilon}_{1,2,3}$
are the radii of the small semicircles.
This is related by a conformal transformation to the determinant for
the $z$-plane figure given by
\bea
-{1\over2}\ln{\det}_z&\sim&{1\over48}\ln{\epsilon}_1
{\epsilon}_2
{\epsilon}_3\epsilon_4-{1\over2}\ln{\det}_{\rm DNDN}
\eea
Since $\partial_n\Sigma=0$ on all of the straight line segments of the
boundary of the rectangle, we only get a contribution from the
change formula near each of the corners. So we approximate $\Sigma$
for each corner in turn. Starting with the upper
left corner, put $z=-K+iK^\prime+re^{i\varphi}$ with $r$ small.
then
\bea
{\rm sn}(-K+iK^\prime)&=& -{1\over k},\quad
{\rm sn}^\prime(-K+iK^\prime)=0,\quad {\rm sn}^{\prime\prime}(-K+iK^\prime)
=-{1-k^2\over k}\\
{\rm sn}(z)&\approx&-{1\over k}-{1-k^2\over 2k}r^2e^{2i\varphi},\quad
u\approx-{(k-1)^2\over4}r^2e^{2i\varphi},\quad 
{\bar\epsilon}_1={(k-1)^2\epsilon_1^2\over4}\\
\Sigma&\approx&\ln{(k-1)^2r\over2},\qquad \partial_n\Sigma=-{1\over r}\\
\Delta_1&=&\left(-{1\over48}-{1\over24}\right)\Sigma=-{1\over16}
\ln{(k-1)^2\epsilon_1\over2}
\eea
For the lower left corner $z=-K+re^{i\varphi}$
\bea
{\rm sn}(-K)&=& -1,\qquad
{\rm sn}^\prime(-K)=0,\qquad {\rm sn}^{\prime\prime}(-K)
={1-k^2}\\
{\rm sn}(z)&\approx&-{1}+{1-k^2\over 2}r^2e^{2i\varphi},\quad
u\approx{(k-1)^2\over(k+1)^2}[1+kr^2e^{2i\varphi}],\quad 
{\bar\epsilon}_2={k(k-1)^2\epsilon_2^2\over(k+1)^2}\\
\Sigma&\approx&\ln{2k(k-1)^2r\over(k+1)^2},\qquad \partial_n
\Sigma=-{1\over r}\\
\Delta_2&=&\left(-{1\over48}-{1\over24}\right)\Sigma=-{1\over16}
\ln{2k(k-1)^2\epsilon_2\over(k+1)^2}
\eea
For the lower right corner $z=K+re^{i\varphi}$,
\bea
{\rm sn}(K)&=& 1,\qquad
{\rm sn}^\prime(K)=0,\qquad {\rm sn}^{\prime\prime}(K)
=-({1-k^2})\\
{\rm sn}(z)&\approx&{1}-{1-k^2\over 2}r^2e^{2i\varphi},\quad
u\approx 1-kr^2e^{2i\varphi},\quad 
{\bar\epsilon}_3={k\epsilon_3^2}\\
\Sigma&\approx&\ln{2kr},\qquad \partial_n
\Sigma=-{1\over r}\\
\Delta_3&=&\left(-{1\over48}-{1\over24}\right)\Sigma=-{1\over16}
\ln{2k\epsilon_3}
\eea
For the final (upper right) corner, $z=K+iK^\prime+re^{i\varphi}$,
\bea
{\rm sn}(K+iK^\prime)&=& {1\over k},\quad
{\rm sn}^\prime(-K+iK^\prime)=0,\quad {\rm sn}^{\prime\prime}(K+iK^\prime)
={1-k^2\over k}\\
{\rm sn}(z)&\approx&{1\over k}+{1-k^2\over 2k}r^2e^{2i\varphi},\quad
u\approx-{4\over(k+1)^2r^2e^{2i\varphi}},\quad 
R={4\over(k+1)^2\epsilon_4^2}\\
\Sigma&\approx&\ln{8\over(1+k)^2r^3},\qquad \partial_n\Sigma={3\over r}\\
\Delta_4&=&\left({3\over48}-{1\over24}\right)\Sigma={1\over48}
\ln{8\over(1+k)^2\epsilon_4^3}
\eea
Then we have
\bea
-{1\over2}\ln{\det}_u&=&-{1\over2}\ln{\det}_z+\Delta_1+\Delta_2+\Delta_3
+\Delta_4\\
&\sim&-{1\over24}\ln{\epsilon}_1
{\epsilon}_2
{\epsilon}_3\epsilon_4-{1\over2}\ln{\det}_{\rm DNDN}-{1\over16}
\ln{2k^2(k-1)^4\over(k+1)^2}+{1\over48}
\ln{8\over(1+k)^2}\nonumber\\
&\sim&-{1\over24}\ln{\epsilon}_1
{\epsilon}_2
{\epsilon}_3\epsilon_4-{1\over2}\ln{\det}_{\rm DNDN}-{1\over16}
\ln{k^2(k-1)^4}+{1\over24}
\ln{(1+k)^2}
\eea
On the other hand
\bea
{1\over48}\ln R-{1\over48}\ln{\bar\epsilon}_1
{\bar\epsilon}_2
{\bar\epsilon}_3&\sim&{1\over48}\ln{4\over(k+1)^2\epsilon_4^2}
-{1\over48}\ln{k^2(k-1)^4\epsilon_1^2\epsilon_2^2\epsilon_3^2\over4(k+1)^2}
\nonumber\\
&\sim&
-{1\over24}\ln{k(k-1)^2\epsilon_1\epsilon_2\epsilon_3\epsilon_4\over4}
\eea
So comparing we deduce
\bea
\ln{\cal D}&=&{1\over24}\ln{k(k-1)^2\over4}-{1\over2}\ln{\det}_{\rm DNDN}-{1\over8}
\ln{k(k-1)^2}+{1\over24}\ln{(1+k)^2}\nonumber\\
&=&-{1\over2}\ln{\det}_{\rm DNDN}-{1\over12}
\ln{2k(k-1)^2}+{1\over24}\ln{(1+k)^2}\nonumber\\
&=&-{1\over2}\ln{\det}_{\rm DNDN}-{1\over24}
\ln{4k^2(k-1)^4\over(1+k)^2}
\label{cald}
\eea
It is important to bear in mind that this formula applies only when
the corners of the rectangle are mapped to $0,U=(1-k)^2/(1+k)^2,1,\infty$,
which mark the DN transition points. The formula to use when
the transition points are at general locations, can be obtained
by executing a projective conformal transformation
\bea
w&=&{au+b\over cu+d},\qquad ad-bc=1.
\eea
Carefully transforming the corresponding determinants,
regulated by suitable circular arcs to avoid singular points, 
leads to the result
\bea
\ln{\cal D}_w=\ln{\cal D}+{1\over8}\ln cd(cU+d)(c+d)
\eea
For example, a symmetrical and canonical choice is to map the
corners of the rectangle to $-1/k,-1,+1,+1/k$ respectively.
For this case, $ad(c+d)(cU+d)=k^2/(1+k)^2$, and
The corresponding determinant ${\cal D}_0$ is given by
\bea
\ln{\cal D}_0&=&-{1\over2}\ln{\det}_{\rm DNDN}-{1\over24}
\ln{4k^2(k-1)^4\over(1+k)^2}+{1\over8}\ln{k^2\over(1+k)^2}\nonumber\\
&=&-{1\over2}\ln{\det}_{\rm DNDN}+{1\over12}\ln{k^2\over2(1-k^2)^2}
\eea
\subsubsection{{\cal D} for Unmixed Boundary conditions}
As we have noted, for unmixed boundary conditions the analog
of ${\cal D}$ had nothing to depend on, and so had to be a constant.
It is instructive to see this using the methods of the present
subsection. 
The determinant for the $u$-plane figure changes, in the unmixed case,
to
\bea
-{1\over2}\ln{\det}_u&\sim&{5\over24}\ln R+{1\over24}\ln{\bar\epsilon}_1
{\bar\epsilon}_2
{\bar\epsilon}_3+\ln{\cal D}^{\rm N}\nonumber\\
&\sim&{1\over12}\ln {{\epsilon}_1
{\epsilon}_2
{\epsilon}_3\over\epsilon^5_4}
+{1\over12}\ln{k(k-1)^2\over(k+1)^6}+\ln{\cal D}^{\rm N}.
\eea
And the determinant for the $z$-plane figure becomes
\bea
-{1\over2}\ln{\det}_z&\sim&{1\over48}\ln{\epsilon}_1
{\epsilon}_2
{\epsilon}_3\epsilon_4-{1\over2}\ln{\det}_{\rm DDDD}
\eea
To relate these we need to adapt the $\Delta_i$ to the unmixed case.
The only difference is that the corner contributions for each
quarter circle add instead of cancel:
\bea
\Delta_1&=&\left(-{1\over48}-{1\over24}+{1\over8}\right)\Sigma=
{1\over16}\ln{(k-1)^2\epsilon_1\over2}\nonumber\\
\Delta_2&=&{1\over16}
\ln{2k(k-1)^2\epsilon_2\over(k+1)^2},\qquad\Delta_3={1\over16}
\ln{2k\epsilon_3}\nonumber\\
\Delta_4&=&\left({3\over48}-{1\over24}+{1\over8}\right)\Sigma={7\over48}
\ln{8\over(1+k)^2\epsilon_4^3}\nonumber\\
\Delta&=&\sum_i\Delta_i={1\over16}\ln{2k^2(k-1)^4
\epsilon_1\epsilon_2\epsilon_3\over(k+1)^2}+{7\over48}
\ln{8\over(1+k)^2\epsilon_4^3}
\eea
Then
\bea
\Delta-{1\over2}\ln{\det}_z&\sim&{1\over12}
\ln{{\epsilon}_1{\epsilon}_2{\epsilon}_3\over\epsilon^5_4}
-{1\over2}\ln{\det}_{\rm DDDD}+{1\over16}\ln{2k^2(k-1)^4
\over(k+1)^2}+{7\over48}
\ln{8\over(1+k)^2}
\eea
Since this quantity should be the determinant in the $u$-plane, we
must have
\bea
\ln{\cal D}^{\rm N}
&=&
-{1\over2}\ln{\det}_{\rm DDDD}+{1\over16}\ln{2k^2(k-1)^4
\over(k+1)^2}+{7\over48}
\ln{8\over(1+k)^2}-{1\over12}\ln{k(k-1)^2\over(k+1)^6}\nonumber\\
&=&-{1\over2}\ln{\det}_{\rm DDDD}+{1\over48}\ln{k^2(k^2-1)^4}+{1\over2}\ln2
\eea
To see that the right side is a constant we use
\bea
k^{1/24}(1-k^2)^{1/12}&=&{\theta_2(0)^{1/12}\theta_4(0)^{1/3}
\over\theta_3(0)^{5/12}}={(\theta_2(0)\theta_3(0)\theta_4(0))^{1/12}
\theta_4(0)^{1/4}
\over\theta_3(0)^{1/2}}\nonumber\\
&=&2^{1/12}q^{1/48}{\prod(1-q^{2n-1})^{1/2}
\over\prod(1+q^{2n-1})}
=2^{1/12}q^{1/48}{\prod(1-q^{n})^{1/2}
\over\sqrt{\theta_3(0)}}\nonumber\\
&=&2^{1/12}q^{1/48}{\prod(1-q^{n})^{1/2}
\over(2K/\pi)^{1/4}}=\pi^{1/4}2^{1/12}{\det}^{+1/2}_{\rm DDDD}\nonumber\\
\ln{\cal D}^{\rm N}&=&{1\over12}\ln(2\pi^3)+{1\over2}\ln2={1\over12}\ln2^7\pi^3
\eea
In a similar vein, executing a projective conformal transformation shows
that ${\cal D}^N$ is a projective invariant.

\end{appendix}

\end{document}